\documentclass[twocolumn]{aastex631}
\usepackage{latexsym}
\usepackage{amssymb}
\usepackage{amsfonts}
\usepackage{amsmath}
\usepackage{bm}
\usepackage[normalem]{ulem}
\usepackage{multirow}
\usepackage{natbib} 
\usepackage{gensymb}
\usepackage{subfigure}
\usepackage{wrapfig}
\usepackage{color}
\usepackage{calc}
\usepackage{amsmath,amssymb,graphicx}
\usepackage{amssymb,amsmath}
\usepackage{tensor}
\usepackage{bm}
\usepackage{pifont}
\usepackage{microtype}
\usepackage{booktabs}
\usepackage{times}
\usepackage[varg]{txfonts}
\usepackage{subfigure}
\usepackage{lipsum}
\usepackage{breakurl} 
\usepackage{graphicx}
\usepackage{pbox}
\usepackage{xspace}
\usepackage{amssymb}

\definecolor{LinkColor}{rgb}{0.75, 0, 0}
\definecolor{CiteColor}{rgb}{0, 0.5, 0.5}
\definecolor{UrlColor}{rgb}{0, 0, 0.75}
\hypersetup{linkcolor=LinkColor}
\hypersetup{citecolor=CiteColor}
\hypersetup{urlcolor=UrlColor}
\allowdisplaybreaks
\DeclareFontFamily{OT1}{pzc}{}
\DeclareFontShape{OT1}{pzc}{m}{it}{<-> s * [1.10] pzcmi7t}{}
\DeclareMathAlphabet{\mathpzc}{OT1}{pzc}{m}{it}

\def\at{AT2019wxt\xspace}

\shorttitle{Multi-wavelength follow-up of AT2019wxt }
\shortauthors{}
\graphicspath{{./}{figures/}}

\begin{document}

\title{AT2019wxt: An ultra-stripped supernova candidate discovered in electromagnetic follow-up of a gravitational wave trigger}



\author[0000-0001-9289-0570]{Hinna Shivkumar}
\affiliation{Anton Pannekoek Institute for Astronomy, University of Amsterdam, Science Park 904, 1098 XH, Amsterdam, The Netherlands}

\author[0000-0002-3850-6651]{Amruta D. Jaodand}
\affiliation{California Institute of Technology, 1200 E California Blvd., Pasadena, CA 91125, USA}

\author[0000-0003-0477-7645]{Arvind Balasubramanian}
\affiliation{Department of Physics and Astronomy, Texas Tech University, Box 1051, Lubbock, TX 79409-1051, USA}

\author[0000-0002-4223-103X]{Christoffer Fremling}
\affiliation{California Institute of Technology, 1200 E California Blvd., Pasadena, CA 91125, USA}

\author[0000-0001-8104-3536]{Alessandra Corsi}
\affiliation{Department of Physics and Astronomy, Texas Tech University, Box 1051, Lubbock, TX 79409-1051, USA}

\author[0000-0003-0484-3331]{Anastasios Tzanidakis}
\affiliation{University of Washington, 3910 15th Avenue NE, Seattle, WA 98195, USA}

\author[0000-0001-6573-7773]{Samaya Nissanke}
\affiliation{GRAPPA, Anton Pannekoek Institute for Astronomy and Institute of High-Energy Physics,\\
University of Amsterdam, Science Park 904,1098 XH Amsterdam, The Netherlands}
\affiliation{Nikhef, Science Park 105, 1098 XG Amsterdam, The Netherlands}

\author[0000-0002-5619-4938]{Mansi Kasliwal}
\affiliation{California Institute of Technology, 1200 E California Blvd., Pasadena, CA 91125, USA}

\author[0000-0002-8147-2602]{Murray Brightman}
\affiliation{California Institute of Technology, 1200 E Califronia Blvd., Pasadena, CA 91125, USA}

\author[0000-0002-9397-786X]{Geert Raaijmakers}
\affiliation{GRAPPA, Anton Pannekoek Institute for Astronomy and Institute of High-Energy Physics,\\
University of Amsterdam, Science Park 904,1098 XH Amsterdam, The Netherlands}

\author[0000-0003-1252-4891]{Kristin Kruse Madsen}
\affiliation{CRESST and X-ray Astrophysics Laboratory, NASA Goddard Space Flight Center, Greenbelt, MD 20771 USA}

\author{Fiona Harrison}
\affiliation{California Institute of Technology, 1200 E California Blvd., Pasadena, CA 91125, USA}

\author[0000-0002-6575-4642]{Dario Carbone}
\affiliation{University of the Virgin Islands, 2 Brewers Bay Road, Charlotte Amalie, USVI 00802, USA}

\author[0000-0002-8070-5400]{Nayana A.J.}
\affiliation{Indian Institute of Astrophysics, II Block, Koramangala, Bangalore 560034, India.}

\author[0000-0002-0875-8401]{Jean-Michel Désert}
\affiliation{Anton Pannekoek Institute for Astronomy, University of Amsterdam, Science Park 904, 1098 XH, Amsterdam, The Netherlands}

\author[0000-0002-8977-1498]{Igor Andreoni}
\affil{Joint Space-Science Institute, University of Maryland, College Park, MD 20742, USA}
\affil{Department of Astronomy, University of Maryland, College Park, MD 20742, USA}
\affil{Astrophysics Science Division, NASA Goddard Space Flight Center, Mail Code 661, Greenbelt, MD 20771, USA}

\correspondingauthor{Hinna Shivkumar, Amruta Jaodand}
\email{h.shivkumar@uva.nl, ajaodand@caltech.edu}


\begin{abstract}
We present optical, radio and X-ray observations of a rapidly-evolving transient AT2019wxt (PS19hgw), discovered during the search for an electromagnetic (EM) counterpart to the gravitational-wave (GW) trigger S191213g \citep{s191213g}. Although S191213g was not confirmed as a significant GW event in the off-line analysis of LIGO-Virgo data, \at remained an interesting transient due its peculiar nature. The optical/NIR light curve of \at displayed a double-peaked structure evolving rapidly in a manner analogous to currently know ultra-stripped supernovae (USSNe) candidates. This double-peaked structure suggests presence of an extended envelope around the progenitor, best modelled with two-components: i) early-time shock-cooling emission and ii) late-time radioactive $^{56}$Ni decay. We constrain the ejecta mass of \at at $M_{ej} \approx{0.20 M_{\odot}}$ which indicates a significantly stripped progenitor that was possibly in a binary system. We also followed-up \at with long-term \textit{Chandra} and Jansky Very Large Array  observations spanning $\sim$260 days. We detected no definitive counterparts at the location of \at in these long-term X-ray and radio observational campaigns. We establish the X-ray upper limit at $9.93\times10^{-17}$\,erg~cm$^{-2}$~s$^{-1}$ and detect an excess radio emission from the region of \at. However, there is little evidence for SN1993J- or GW170817-like variability of the radio flux over the course of our observations. A substantial host galaxy contribution to the measured radio flux is likely. The discovery and early-time peak capture of \at in optical/NIR observation during EMGW follow-up observations highlights the need of dedicated early, multi-band photometric observations to identify USSNe. 
\end{abstract}





\section{Introduction}
\label{sec:Intro}
Massive stars at the endpoints of their lives undergo mass loss through ejection of some or all of their hydrogen (and possibly helium) envelopes, eventually collapsing in what are known as stripped-envelope core-collapse supernovae \citep[SESNe,][]{Filippenko:1997, snibc}. The extent to which the outer layers of massive stars are stripped dictates their spectroscopic classification into their various sub-classes. Partial stripping in Type IIb SNe is supported by the presence of Balmer lines while strong stripping in Type Ic SNe is evident by absence of both Hydrogen and Helium lines. The current population of SESNe suggests that ejection of the progenitor envelopes can be driven by a) mass loss via stellar-wind \citep{BS:1986, WW:1995, Pod:2001}, or b) mass transfer during binary interaction \citep{Pod:1992, snbinary1, snbinary2, Yoon:2017}. 

Large uncertainties currently persist in our understanding of the progenitors of SESNe. Specifically, if any links exist between the various SESNe sub-classes, and if different sub-classes have preferred mass loss mechanisms. At the same time, the observational picture of SESNe has been evolving in the last few years as wide-field optical surveys have accelerated the rate of discovery and started filling-up the luminosity ``gap" between novae and SNe in the luminosity-duration phase-space \citep{lum_time}. In particular, optimized follow-up observations have opened up the parameter space for characterisation of rapidly-evolving, low-luminosity (L$_{bol}=10^{42}$\, erg/s) transients -- resulting in a growing population of ultra-stripped SNe \citep[USSNe][]{KKG:2010,DSM:2013,DKC:2018, JMK:2020, de_gqr}. 

The progenitors of USSNe undergo extreme envelope stripping via two stages of common envelope evolution, with a first phase of mass-transfer through Roche-Lobe overflow, leading to a He-star NS system. The second mass-transfer phase in the resulting He-star NS system involves stripping of the He-star leading to a stripped He-star NS system with a He-rich envelope. The core-collapse of the stripped He star triggers a maximally stripped SN explosion known as USSN. This explosion is accompanied by ejection of $\approx0.01 M_{\odot}$ of the star mass. Such USSN are then expected to evolve into binary neutron star (BNS) systems \citep{tauris2013, tauris2015, tauris2017}. In fact, light curves of these SNe display double peaks in both bluer and redder bands, indicating presence of an extended envelope around the progenitor \citep{nakarpiro}. The combination of shock-cooling emission (SCE) and radioactive decay of $^{56}$Ni has been used to explain this double-peaked structure and rapid evolution in the light curves of currently known USSNe candidates such as SN2019dge \citep{yao_dge} and iPTF14gqr \citep{de_gqr}.

In this work, we present multi-wavelength observations spanning optical, radio and X-ray wavebands for one such puzzling USSN candidate dubbed \at. \at was discovered by the PanSTARRS Search for Kilonovae survey (as PS19hgw) during their EM follow-up of LIGO-Virgo GW trigger, S191213g (flagged as a BNS merger event; \citealp{s191213g}). A follow-up observational campaign across the EM spectrum was encouraged because \at was located in the 80\% confidence contour of the S191213g's skymap \citep{gcnskyloc} with its distance being consistent with estimated luminosity distance range for S191213g \citep{McBrien:2019}. Moreover, the optical light curve of \at displayed a very fast decline in comparison to previously known rapidly-evolviong, hydrogen-free supernovae such as SN2008ha or SN2010ae (\citet{valenti2009}; \citet{foley2009}; \citet{stritzinger2014}). Although, in offline analysis of GW data S191213g was demoted as a significant GW candidate \citep{ligo:O3b}, AT2019wxt remains a very interesting transient. Our multi-band optical observations show a double-peaked rapidly-declining light curve resembling that of the USSNe candidates iPTF14gqr \citep{de_gqr} and SN2019dge \citep{yao_dge}.

\begin{figure*}[ht!]
    \begin{center}
        \includegraphics[width=0.85\textwidth]{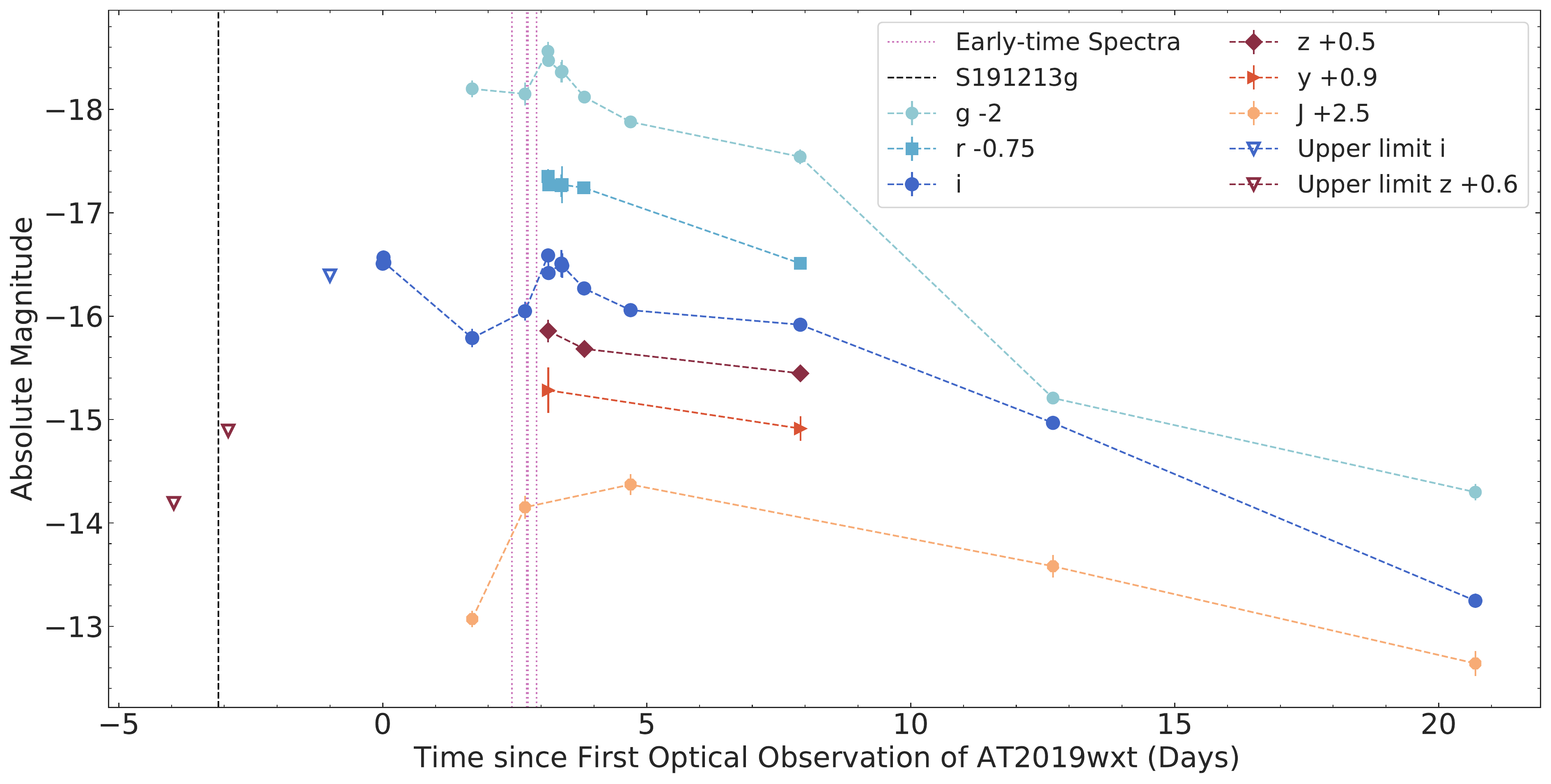}
        \caption{Galactic extinction corrected optical and NIR light curves of the transient AT2019wxt, using the data from Table \ref{tab:table_photometry}. The cyan, light blue, blue, magenta, dark red, orange, and yellow markers represent photometric data in the \textit{g}, \textit{r}, \textit{i}, \textit{z}, \textit{y}, and \textit{J} bands, respectively and the markers for each band are connected by a dashed line to visually track the photometric evolution. The magnitudes are offset vertically for better visibility and the times displayed are relative to the first optical observation at MJD 58833.305, used as a reference epoch henceforth. The vertical black dashed line indicates the time of the GW trigger S191213g, for reference.Vertical, dotted, magenta lines indicate epochs where early-time spectroscopic observations used in this work (see, \ref{sec:spec_obs}).}
        \label{fig:figure_lightcurve}
    \end{center}
\end{figure*}

We outline the interesting nature of \at based on a campaign of high-cadence optical observations and other complementary data from radio and X-ray observational campaigns. Our work is organized as follows. Section \ref{sec:obs} outlines the observational properties of \at and its multi-band follow-up observations. Section \ref{sec:DA} outlines the data analysis techniques used for optical, X-ray and radio observations. Here, we also present a fully Markov-Chain Monte-Carlo (MCMC) parameter estimation for characterization of \at's physical properties. Finally, in section \ref{sec:disc}, we provide interpretation of results obtained from this comprehensive dataset and their implications in the broader context of stellar evolution.

\section{Discovery and panchromatic follow up} \label{sec:obs}
\subsection{AT2019wxt (PS19hgw)}
\at was discovered on 2019 December 16 UTC 07:19:12 (MJD 58833.305) as a source at an optical magnitude of 19.38 $\pm$ 0.05 in the PS1 \textit{i}-band \citep{MSS:2019, McBrien:2019}. It was localised to RA = 01:55:41.941, Dec = +31:25:04.55 and found to be associated with the compact host galaxy, KUG 0152+311\footnote{Galaxy morphology and classification from NASA/IPAC Extragalactic Database (NED) \url{http://ned.ipac.caltech.edu}. Please also refer to \citet{GG:2020}}. The association of \at with KUG 0152+311 was confirmed by early optical spectra \citep{gcnspectra1}, which displayed standard galaxy lines at a redshift of z=0.036 (luminosity distance, d = 144 Mpc). This distance estimate from optical observations placed AT2019wxt within 80\% confidence interval of LIGO-VIRGO localisation skymap \citep[at a luminosity distance range of S191213g $d_L$=201$\pm$81\,Mpc;][]{McBrien:2019,gcnskyloc}. The offset between \at and the host galaxy was observed to be 0.5''S, 7.7'' E with a projected a distance of $6.7$\,kpc from the galactic center \citep{MSS:2019}.

The presence of Helium in optical spectra and rapid photometric evolution further confirmed AT2019wxt as an interesting EM candidate counterpart of S191213g. The rapid decline in brightness was observed to be faster than that for known hydrogen-free SNe \citep[e.g. SN2008ha, SN2010ae][]{valenti2009,foley2009,stritzinger2014} but slower than the kilonova AT2017gfo associated with GW170817 (e.g. \citealp{gfo3}; \citealp{cowper_kn}) and the possible white dwarf-neutron star (WD-NS) merger SN2018kzr \citep{2018kzr,gcnsmartt}. Hence, we triggered a comprehensive, multi-band EM follow-up campaign with a host of space and ground-based telescopes. This observational campaign spanned optical/near-Infrared (NIR), X-ray, and radio wavebands. Observations for each of these wavebands are summarized in Tables \ref{tab:table_photometry}, \ref{tab:specs}, \ref{tab:x_ray}, and \ref{tab:radio_obs}, respectively.

\subsection{Optical/NIR Observations}
\subsubsection{Photometric Observations}
After the initial discovery of \at was reported by PanSTARRS \citep{MSS:2019, McBrien:2019}, the Global Relay of Observatories Watching Transients Happen (GROWTH) collaboration conducted further follow-up observations using the Spectral Energy Distribution Machine \citep[SEDM;][]{bnw:2018} on the Palomar 60-inch \citep[P60][]{BBM:2006} telescope. The SEDM obtained 180\,s exposure time images of \at with the rainbow camera imager for each of the  \textit{ugri} filters. These images were processed using a standard python-based and fully automated reduction pipeline FPipe \citep{fpipe:2016}, which performs host-galaxy subtraction and PSF fitting photometry. Host-galaxy subtraction was performed using SDSS images of KUG 0152$+$311, and the source photometry was derived in the AB magnitude system \citep{gcnfrem}.

The GROWTH collaboration also obtained 300\,s exposure images in \textit{g}, \textit{r} and \textit{i} filters with the Lulin 1-m Telescope (LOT) located in Taiwan. The LOT magnitudes, also in the AB magnitude system, are calibrated against the PS1 catalog \citep{gcnlulin}. Follow-up observations of \at were also conducted with the Large Monolithic Imager (LMI) \citep{LMI:2014}  on the 4.3m Lowell's Discovery Channel Telescope (DCT, located in Arizona) for each of the \textit{griz} filters. The magnitudes are calibrated with the SDSS catalog and are presented in the AB system \citep{gcndct}. Simultaneously, optical observations of \at were also undertaken with the three channel imager 3KK camera \citep{LBG:2016} on 2m telescope at the Wendelstein Observatory. Observations were obtained on 5 epochs for each of the filters (\textit{g',i',J}). Aperture photometry was performed using eight comparison stars within the field of view of the detector. Magnitude errors include statistical error in the measurement of the magnitude of \at and in the zero-point calculation \citep{gcnwend}. 

The observations and photometric measurements are summarized in Table \ref{tab:table_photometry} and span $\approx{20.7}$ days since initial detection. The multi-band light curves are collectively displayed in Figure \ref{fig:figure_lightcurve}. We corrected apparent magnitudes for Galactic extinction, using the data available on the foreground galactic extinction for the host galaxy KUG 0152+311 on the NASA/IPAC Extragalactic Database (NED) for each band. The NED calculates Galactic extinction values assuming the \citet{galacticextinction} reddening law with $R_{V} \equiv A(V)/E(B-V) = 3.1$.

\subsubsection{Spectroscopic Observations}\label{sec:spec_obs}
Early-time spectroscopic observations of \at were taken on 2019 December 18 and 19 (see, Table \ref{tab:specs}). The initial spectroscopic observations were unable to firmly classify the transient \citep{gcnspectra1,gcnspectra4,gcnspectra5}. AT2019wxt showed narrow lines consistent with the host galaxy redshift of z=0.037, and a blue, relatively featureless continuum with a broad feature at 5400 - 6200\,\AA. \citet{gcnspectra3} identified the broad feature as HeI lines, and suggested that AT2019wxt was either a young Type Ib or perhaps Type IIb supernova given the blue continuum. The similarities of the spectra to SN\,2011fu \citep{SN2011fu} prompted \citet{gcnspectra6} to classify AT2019wxt as a type IIb. This supernova classification was subsequently supported by \citet{gcnspectra7} and \citet{gcnspectra8}.

In this work we use early-time spectroscopic observations obtained with various telescopes such as: i) 2\,m Himalayan Chandra Telescope (HCT) at Indian Astronomical Observatory at Hanle, ii) 8.2\,m Very Large Telescope (VLT) UT1 at European Southern Observatory (ESO) at La Silla, Chile, iii) 3.58\,m New Technology Telescope (NTT) as part of the ePESSTO extended-Public ESO Spectroscopic Survey for Transient Objects (PI:Smartt), and iv) 8.4\,m The Large Binocular Telescope (LBT) at LBT Observatory in Arizona, USA. The HCT observations were conducted with Hanle Faint Object Spectrograph Camera (HFOSC2) instrument \citep{gcnspectra1}. HCT/HFOSC2 provides low to medium resolution grism spectroscopy with resolution of 150 to 4500 based on grism settings. HCT observations in this work used grism 7 was to provide a resolution of 1200 for the observations in the wavelength range of 3800-7500\,\AA. The VLT observations were carried out with FOcal Reducer/low dispersion Spectrograph 2 (FOS2) instrument on UT1 Cassegrain focus in long-slit mode (slit-width of 418.57") to obtain spectroscopic observations of \at \citep{gcnspectra3}. The long-slit mode provides resolution of 260-2600. ESO-NTT was used with ESO Faint Object Spectrograph (EFOS) on Nasmyth B focus \citep{gcnspectra2}. EFOS provides low-resolution spectroscopy of faint objects. The LBT was used with Multi-Object Double Spectrograph which provides a spectral resolution 103 to 104 \citep{gcnspectra6}. 

\subsection{X-ray Observations}
We obtained high resolution ($\sim$1 arcsec) X-ray imaging observations of \at with \textit{Chandra} X-ray. These observations were performed with back-illuminated Advanced CCD Imaging Spectrometer (ACIS) \footnote{for more details please refer to Chapter 6 of \href{https://cxc.harvard.edu/proposer/POG/pdf/MPOG.pdf}{Proposers' Observatory Guide, Cycle 24}} chip S3 in timing exposure (TE) mode. The ACIS S3 chip provided energy resolution of 95\,eV (at 1.49\,keV) at aim-point (in this case, S3 chip), timing resolution of 3.2\,s, and a field-of-view (FOV) of 8.3$\times$8.3\,arcmins. The TE mode allowed for Very Faint (VF) telemetry format which reduced background contamination especially at lower and higher energy ranges for low count rate and/or extended sources.

Our trigger criterion for these \textit{Chandra} observations was a well-localised ($\sim$few arc-seconds) BNS merger within 200\,Mpc. At the time of initial discovery, \at was a primary and interesting candidate counterpart to S191213g given the reddening and photometric evolution. Moreover, the host galaxy redshift firmly placed it within $3\sigma$ (100\,Mpc) of the S191213g candidate (201$\pm$81 Mpc; \citealp{gcnskyloc}). Extrapolating from GW170817 \citep{Haggard:2017}, we expected at least 10 photons in one 100\,ks  exposure. The observations were scheduled to span 6 months period post initial GW trigger (a detailed summary of the observations can be found in Table. \ref{tab:x_ray}). This timescale was motivated by continued X-ray observations of GW170817, 3.5 yrs post the BNS merger event \citep{Hajela:2022}. Given lower total mass prediction for S191213g than GW170817, we expected a long-lived supermassive or stable NS remnant with X-ray emission from the magnetar wind nebula emerging at later times, and hence continued the \at follow-up.
\subsection{Radio Observations}\label{sec:radio_obs}
Radio observations of the \at field were carried out using the Karl G Jansky Very Large Array (VLA) between UT 19 December 2019 and 20 August 2020 in the D (VLA:19A-222, PI:Troja), C and B configurations (VLA:18B-320 and VLA:20A-115, PI: Frail). These observations were performed at nominal central frequencies of 10\,GHz (X-band), 15\,GHz (Ku-band), 22\,GHz (K-band), and are summarised in Table \ref{tab:radio_obs}. The raw data were calibrated using the \texttt{CASA} \citep{2007ASPC..376..127M} automated calibration pipeline and imaging analysis was performed using the \texttt{CASA} task \texttt{tclean}. The relative observational epochs in column two of Table \ref{tab:radio_obs} are with respect to the first optical detection (reference epoch; MJD 58833.305). 

\begin{figure}[!ht]
    \centering
    \includegraphics[width=0.85\columnwidth,trim={0 0 0 1.5cm},clip]{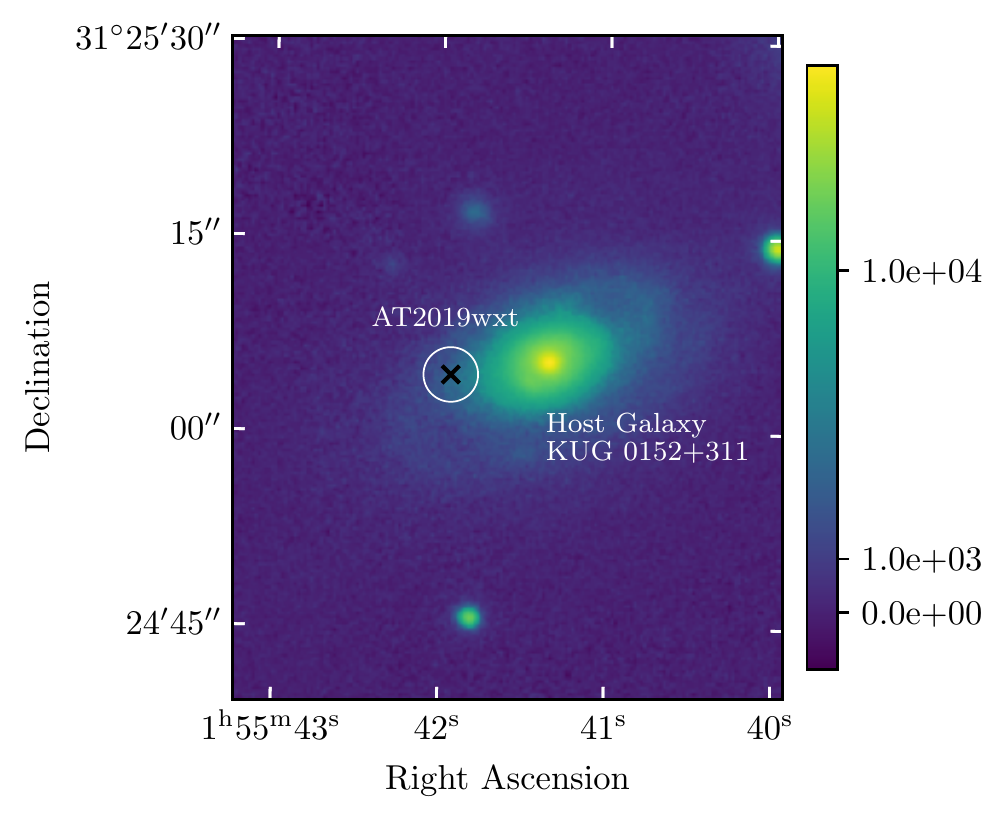}
    \caption{Pan-STARRS i-band image overplotted with the position of \at (cross marks the optical position) and a circle of radius 2.1$^{\prime\prime}$ along with extent of the host galaxy KUG 0152+311.}
    \vspace{-0.35 cm}
    \label{fig:figure_wxtimage}
\end{figure}

In Table \ref{tab:radio_obs} we also present flux densities of AT2019wxt and its host galaxy KUG 0152+311 (in units of $\mu$Jy $=10^{-29}$\,erg\,s$^{-1}$\,cm$^{-2}$\,Hz$^{-1}$). The flux density of AT2019wxt is the maximum value obtained from a circular region of one nominal synthesised beam width\footnote{\url{https://science.nrao.edu/facilities/vla/docs/manuals/oss/performance/resolution}} centered on \at using the \texttt{CASA} task \texttt{imstat}. Using this same task in the residual image, we computed root mean square (RMS) flux density within a 30\,$^{\prime\prime}$ circular region centered on \at. The flux calibrator, 3C48, used in these computations has been undergoing a flare since January 2018\footnote{\url{https://science.nrao.edu/facilities/vla/docs/manuals/oss/performance/fdscale}}. Hence, we added additional 10\% (X and Ku bands) and 20\% (K-band) absolute flux calibration errors in quadrature to the above RMS values. These final flux density errors are presented in the Table \ref{tab:radio_obs}. Any observation with resulting flux density lower than $3\times$ the final flux density error were designated as upper limits. All K-band observations are therefore upper limits.

We then obtained the peak flux density of the host galaxy KUG 0152+311 from \texttt{CASA} task \texttt{imstat} in circular regions of radii 2.1$^{\prime\prime}$ (X-band), 1.4$^{\prime\prime}$ (Ku-band) and 3.1$^{\prime\prime}$ (K-band), centred at 01h55m41.363s, 31h25m05.06s in all configurations (so as to account for extended emission from the host, see Fig. \ref{fig:figure_wxtimage}). Absolute flux calibration errors as described above were added in quadrature to the RMS value obtained from the large 30\,arcsec region around \at to obtain the error on the galaxy's peak flux density.

\section{Data Analysis \& Results}\label{sec:DA}

\subsection{Optical/NIR Data Analysis}
\subsubsection{Light curve Evolution}
The earliest optical detection of \at was obtained in \textit{i}-band and the light curve in this band displayed a prominent double-peaked structure (evident from Figure \ref{fig:figure_lightcurve}). While g-band confirms a similar double-peaked structure, we only observe the tail-end of first peak due to lack of early-time observations. We lack early-time observations during the rise to the second peak in the \textit{r}, \textit{z} and \textit{y} bands, so in these bands we can only report a decline in brightness relative to the time of the second peak in \textit{g} and \textit{i} bands.
 
\begin{figure}[!ht]
    \centering
    \includegraphics[width=0.9\columnwidth]{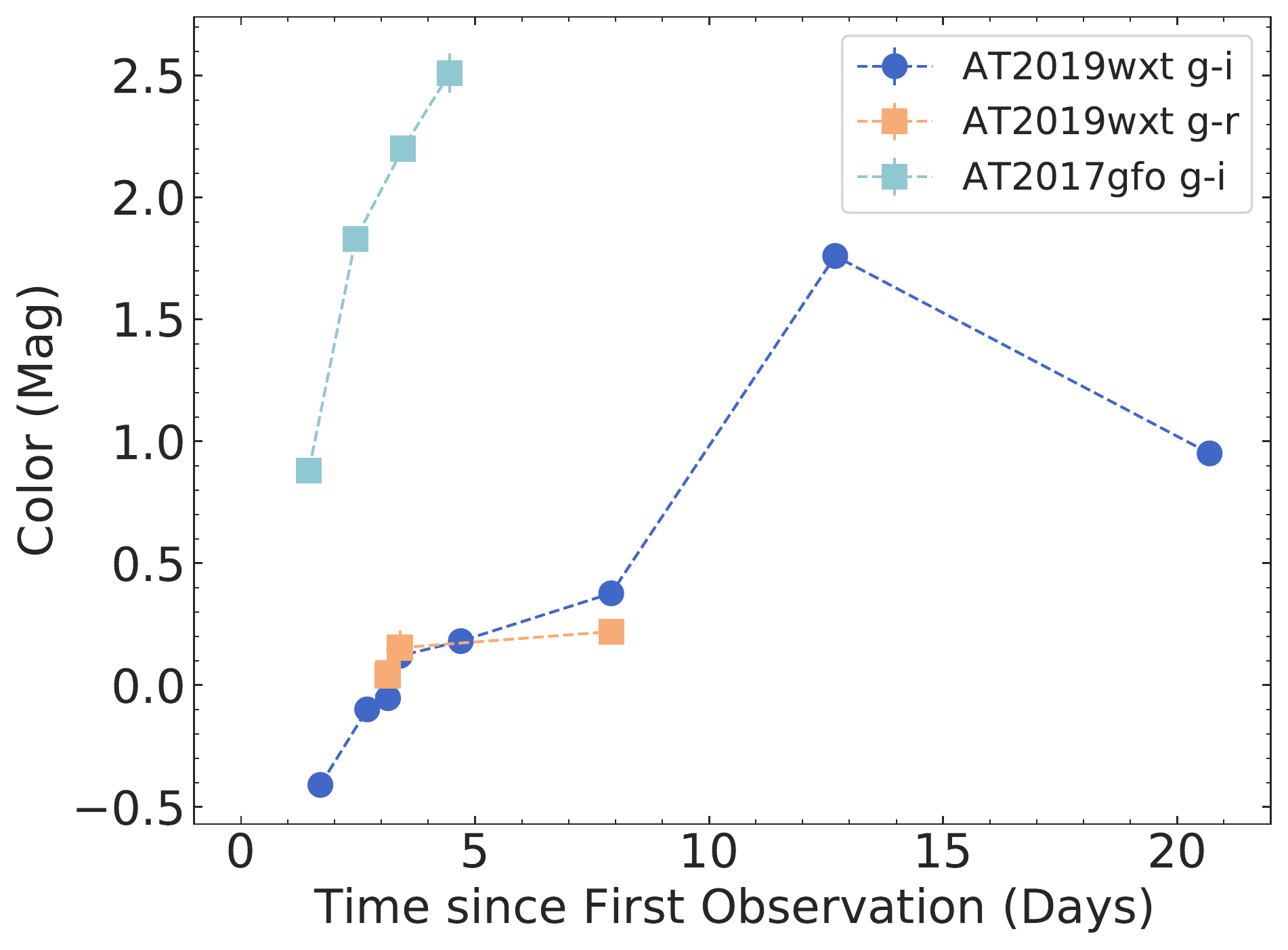}
    \caption{Color evolution of AT2019wxt compared to that of the GW170817 post-merger kilnova, AT2017gfo. Blue solid circles show the \textit{g\textminus\ i} color evolution for \at and orange filled squares show the \textit{g\textminus\ r} color evolution for \at. A reddening through the \textit{g\textminus\ i} color evolution is observed followed by a peak at $\sim$13\,days post initial transient detection. Meanwhile, cyan filled squares and dashed connector lines between them show the \textit{g\textminus\ i} color evolution for AT2017gfo.}
    \label{fig:figure_color}
    \vspace{-0.175 in}
\end{figure}

The rate of evolution of the optical light curve is quite rapid and comparable to the previously known fastest, extra-galactic transients (further discussed in Sec. \ref{sec:disc}). For example, in the \textit{i}-band, the first peak (M$\textsubscript{i, peak1} \approx{-16.5 \pm 0.05}$ mag) displays a rapid decline in brightness, reaching a minimum within $\approx{1.7}$ days. A subsequent rebrightening is then observed which leads to a second peak (M$\textsubscript{i, peak2} \approx{-16.6 \pm 0.07}$ mag) at $\approx{3.1}$ days since initial detection. 
After the second peak, the light curves exhibit a rapid decline in the bluest bands (\textit{g,i}), whereas a relatively shallow ($\approx{0.108}$ mag day$^{-1}$) decline is seen in the redder NIR \textit{J}-band. Although, we note that this shallow decline may just be a result of a lack of observations in the \textit{J}-band. Interestingly, a non-uniform behavior in the light curve evolution is observed post the second peak. It can be characterised by an initial rapid decline up to $\approx{5}$ days at average rate of $\approx{0.43}$ mag day$^{-1}$ in the \textit{g}-band and $\approx{0.27}$ mag day$^{-1}$ in the \textit{i}-band. Following this phase, we observe a decline in the light curve that is relatively slower and accompanied by a shoulder at $\approx{8}$ days. This long-term evolution of the light curve is visible in the \textit{g}, \textit{i} and \textit{J} bands. It should be noted that a similar behaviour with distinct evolution rates between redder and bluer bands has been observed for the kilonova AT2017gfo \citep{cowper_kn}. 

To understand the color evolution of AT2019wxt we calculated \textit{g \textminus \ i} and \textit{g \textminus \ r} colors. To achieve this we initially grouped together observations from different bands performed either simultaneously or on the same day. The colour evolution obtained as such is displayed in Figure \ref{fig:figure_color} and shows \textit{g \textminus \ i} color reddening from 1.7 days to about 6 days at a rate of $\approx{0.13}$ mag day$^{-1}$. This color evolution is slower than that of the kilonova AT2017gfo. This further confirms the distinct and complex evolutionary behavior across different bands.


\subsubsection{Spectral Energy Distribution}
We construct the spectral energy distribution (SED) of AT2019wxt using multi-band photometric observations spanning \textit{grizyJ} bands. 
\begin{figure}[!ht]
    \centering
    \includegraphics[width=0.9\columnwidth]{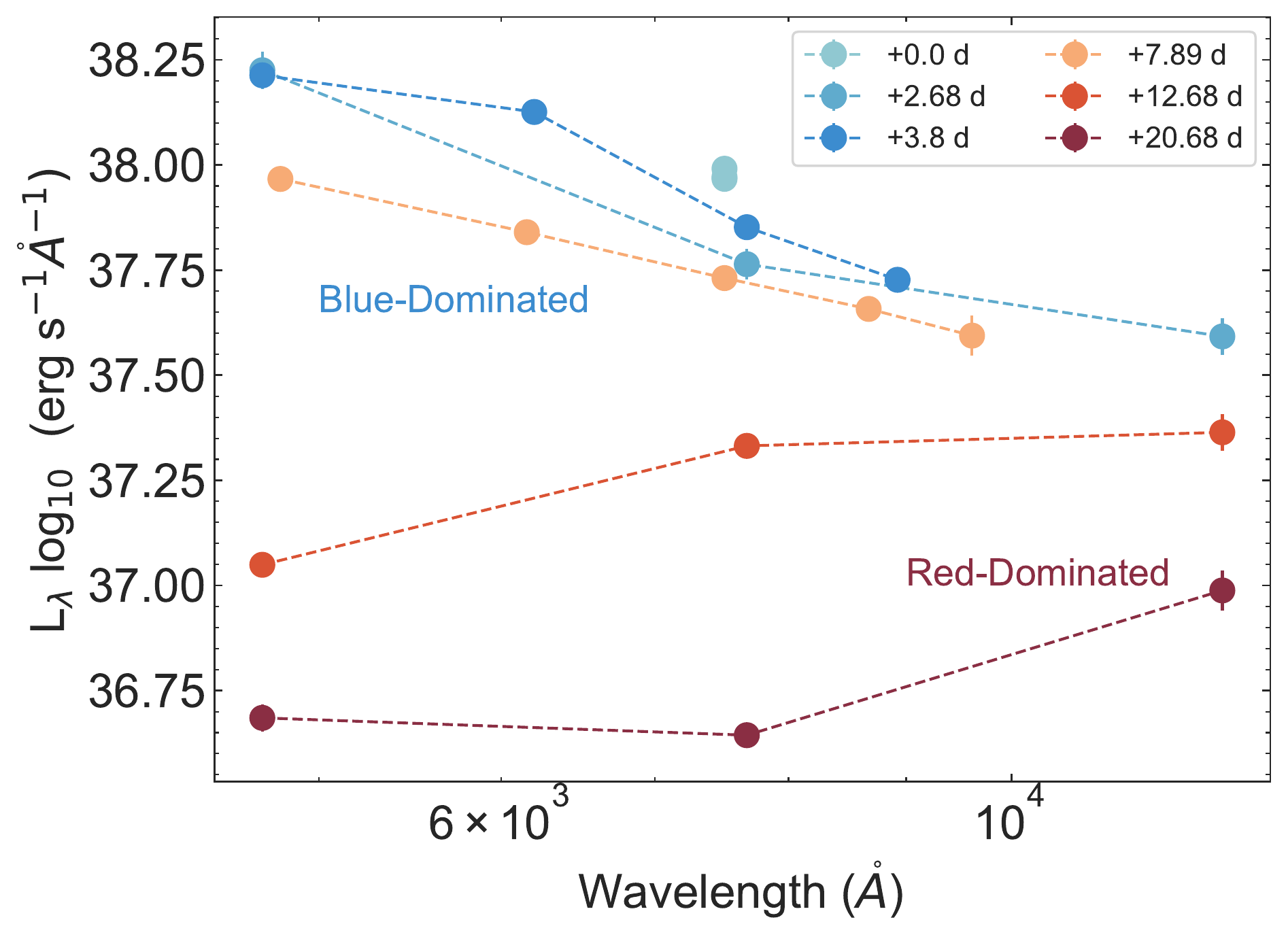}
    \caption{The spectral energy distribution of AT2019wxt at six representative epochs. The cyan, light blue, blue, yellow, dark red and magenta circular markers represent six different epochs, respectively in an increasing order and the markers for each epoch are connected by a dashed line to visually track the evolution of the SED. The early-time emission peaks at bluer wavelengths and we observe a rapid transition to redder wavelengths at late-time. All times are relative to first optical observation as a reference epoch.}
    \label{fig:figure_sed}
\end{figure}
We first grouped these observations from different bands which were performed either simultaneously or on the same day. Of the resulting nine epochs, we present six epochs in Fig. \ref{fig:figure_sed} to compare the temporal evolution of the SED of AT2019wxt. 

\begin{figure*}[!htbp]
    \centering
    \includegraphics[width=0.85\textwidth]{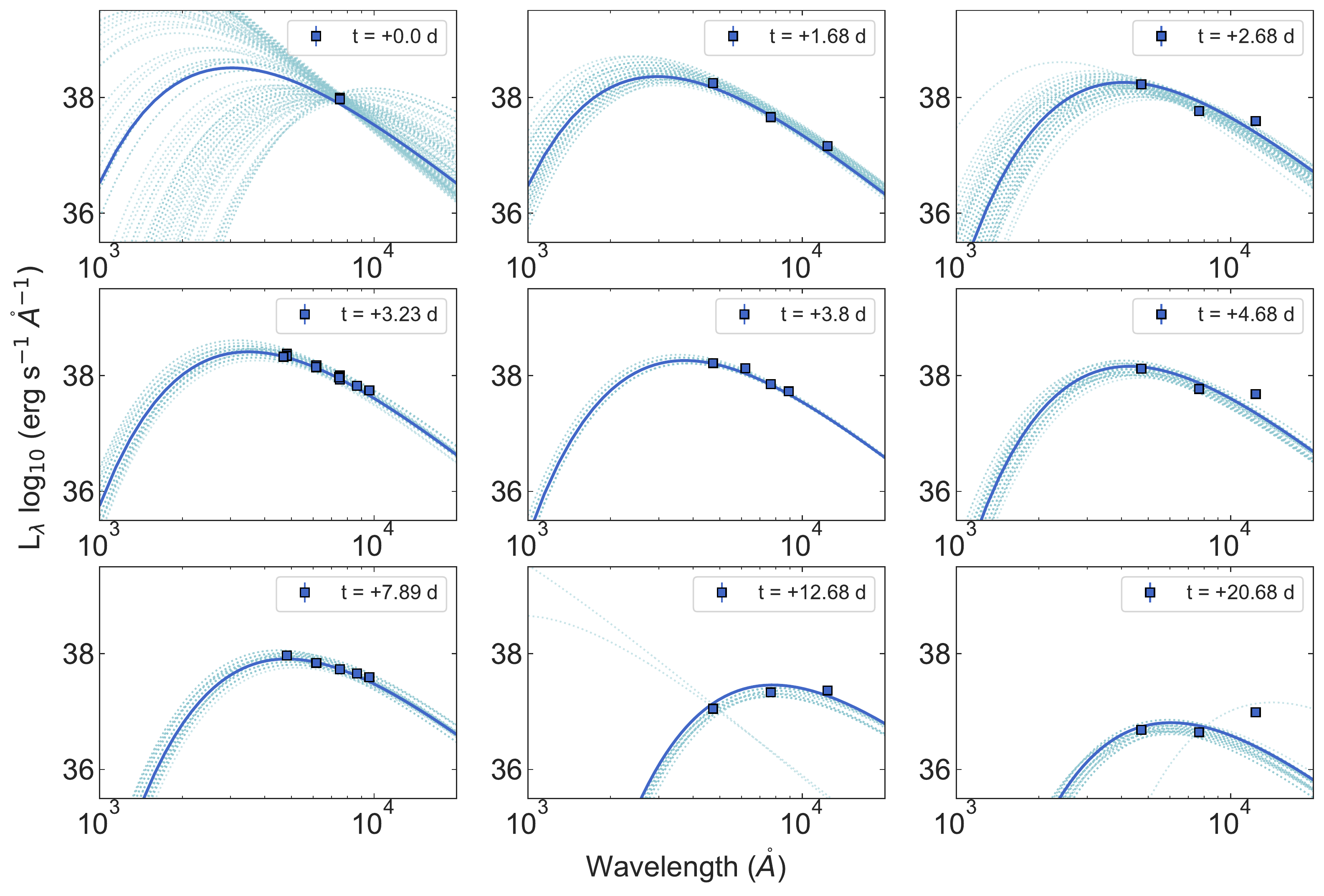}
    \caption{Spectral energy distribution (SED) evolution over eight epochs (days) for AT2019wxt. For each subplot, the square data points indicate the optical photometry at given wavelengths, the solid dark blue line represents the best fit blackbody model from the MCMC inference, and the dotted light blue lines represent the 100 random draws from the MCMC posterior.}
    \vspace{0.5 cm}
    \label{fig:figure_bbmcmc_sed}
\end{figure*}

\begin{figure}[htp]
    \centering
    \includegraphics[width=\columnwidth]{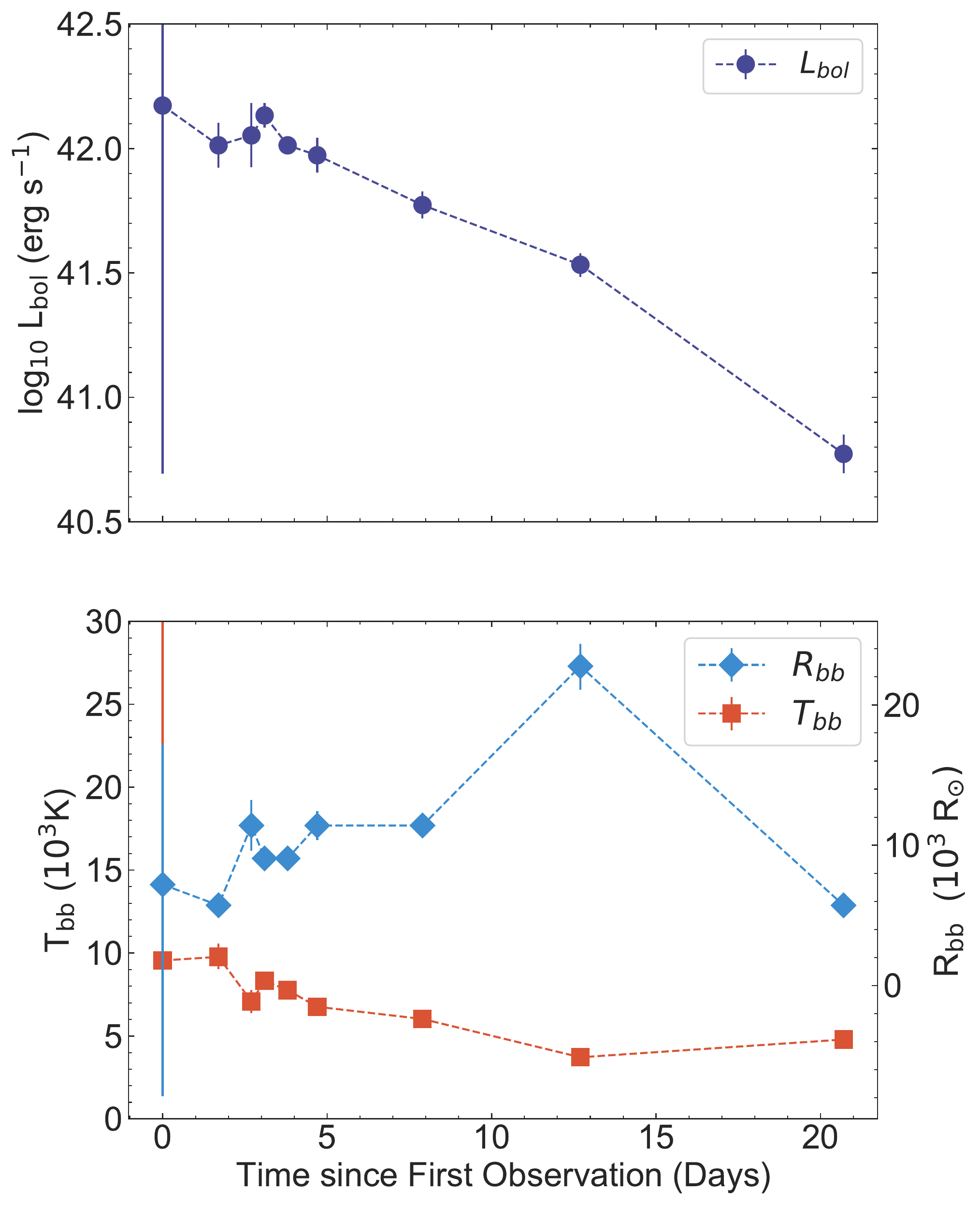}
    \caption{Time evolution of bolometric luminosity, blackbody temperature and blackbody radius obtained from the blackbody fits on the spectral energy distribution using the multi-color photometry for AT2019wxt using PAN-STARRS, Palomar, Lulin, DCT, and Wendelstein 2-m telescopes. All times are relative to the first optical observation as a reference epoch.}
    \label{fig:figure_parms}
    \vspace{-0.5 cm}
\end{figure}

Initially, emission at bluer wavelengths dominates the SED, and over time, we observe a peak shift to redder wavelengths. Assuming the transient emits as a blackbody, as a first-order approximation  we fit a blackbody model to the SED at each epoch using the MCMC implementation via \texttt{emcee} package\footnote{\url{https://github.com/dfm/emcee}} \citep{emcee}. This package uses an MCMC ensemble sampler immune to affine transformations \citep{GW:2010}. Resulting SED and blackbody fits for each epoch are shown in Figure \ref{fig:figure_bbmcmc_sed}. Our MCMC implementation also yielded the blackbody radius (R$\textsubscript{bb}$) and temperature of the blackbody T$\textsubscript{bb}$ at each epoch. The bolometric luminosity,  L$_{bol}$, is then calculated using these parameters (see, Table \ref{tab:table_bbparms}) with a simple Stefan-Boltzmann law. We also show evolution of L$\textsubscript{bol}$, T$\textsubscript{bb}$ and R$\textsubscript{bb}$ in Figure \ref{fig:figure_parms}.

\begin{deluxetable}{ lccc}[!ht]
\renewcommand{\arraystretch}{1.25}
\tablehead{
\colhead{Epoch (MJD)} & \colhead{log$_{10}$L$\textsubscript{bol}$ (erg s$^{-1}$)} &\colhead{R$\textsubscript{bb}$ ($10^{3}$R$_{\odot}$)} & \colhead{T$\textsubscript{bb}$ ($10^{3}$K)}}
\startdata
\hline
58833.3 & 42.29$^{+2.79}_{-1.47}$ & 9.06$^{+12.32}_{-19.12}$ & 9.12$^{+27.95}_{-7.38}$ \\
58835.0 & 42.01$^{+0.09}_{-0.09}$ & 5.72$^{+0.54}_{-0.57}$ & 9.77$^{+0.79}_{-0.73}$ \\
58836.0 & 42.05$^{+0.13}_{-0.13}$ & 11.41$^{+1.81}_{-1.79}$ & 7.08$^{+0.71}_{-0.69}$ \\
58836.4 & 42.13$^{+0.05}_{-0.05}$ & 9.06$^{+0.54}_{-0.54}$ & 8.32$^{+0.32}_{-0.31}$ \\
58837.1 & 42.01$^{+0.02}_{-0.02}$ & 9.06$^{+0.19}_{-0.19}$ & 7.76$^{+0.11}_{-0.11}$ \\
58838.0 & 41.97$^{+0.07}_{-0.07}$ & 11.41$^{+1.02}_{-1.04}$ & 6.76$^{+0.34}_{-0.33}$ \\
58841.2 & 41.77$^{+0.05}_{-0.05}$ & 11.41$^{+0.79}_{-0.79}$ & 6.03$^{+0.23}_{-0.23}$ \\
58846.0 & 41.53$^{+0.05}_{-0.05}$ & 22.76$^{+1.59}_{-1.67}$ & 3.72$^{+0.08}_{-0.08}$ \\
58854.0 & 40.77$^{+0.08}_{-0.08}$ & 5.72$^{+0.63}_{-0.64}$ & 4.79$^{+0.22}_{-0.21}$\\
\hline
\enddata
\caption{\label{tab:table_bbparms} Evolution of bolometric luminosity (L$\textsubscript{bol}$), blackbody radius (R$\textsubscript{bb}$), and blackbody temperature (T$\textsubscript{bb}$) obtained from blackbody fits to the spectral energy distribution of AT2019wxt at different epochs.}
\vspace{-1.5 cm}
\end{deluxetable}

\subsubsection{Bolometric Light Curve Evolution}
The bolometric luminosity of \at peaks at $\approx{1.96}\times 10^{42}$\,erg\,s$^{-1}$ during the first observation. This is followed by a subsequent decline after which a second, lower luminosity peak is observed ($\approx{1.36}\times 10^{42}$\,erg\,s$^{-1}$) at $\approx{3.1}$ days. After this, we observe a shallow decline in the bolometric luminosity. It should be noted that the L$_{bol}$ value estimated from the first observation (via blackbody fits) has large uncertainty due to the lack of multi-band observations. 

The blackbody temperature, T$\textsubscript{bb}$, reaches a maximum value of $\approx{9770}$\,K $\approx{1.5}$ days after the first observation, and rapidly decreases afterwards. At $\approx{3}$ days after the first observation, T$\textsubscript{bb}$ approaches a second peak with a maximum of $\approx{8320}$\,K. On the other hand, the blackbody radius of \at increases over time and reaches a maximum ($\approx{22.76} \times 10^{3}\ R_{\odot}$) at $\approx{12.7}$ days. The inverse trend in T$\textsubscript{bb}$ and R$\textsubscript{bb}$ evolution can be explained by an expanding envelope. Moreover, the high, initial temperature can be attributed to opaque, ionized material. As the matter expands and cools the opacity of this material should decrease with time due to recombination. Here, it is interesting to note that the reddening observed in the \textit{g \textminus \ i} peaks at $\approx{12.7}$ days, which coincides with a maxima in blackbody radius R$\textsubscript{bb}$ and a minima in blackbody temperature. 

\begin{figure}[htp]
    \centering
    \includegraphics[width=\columnwidth]{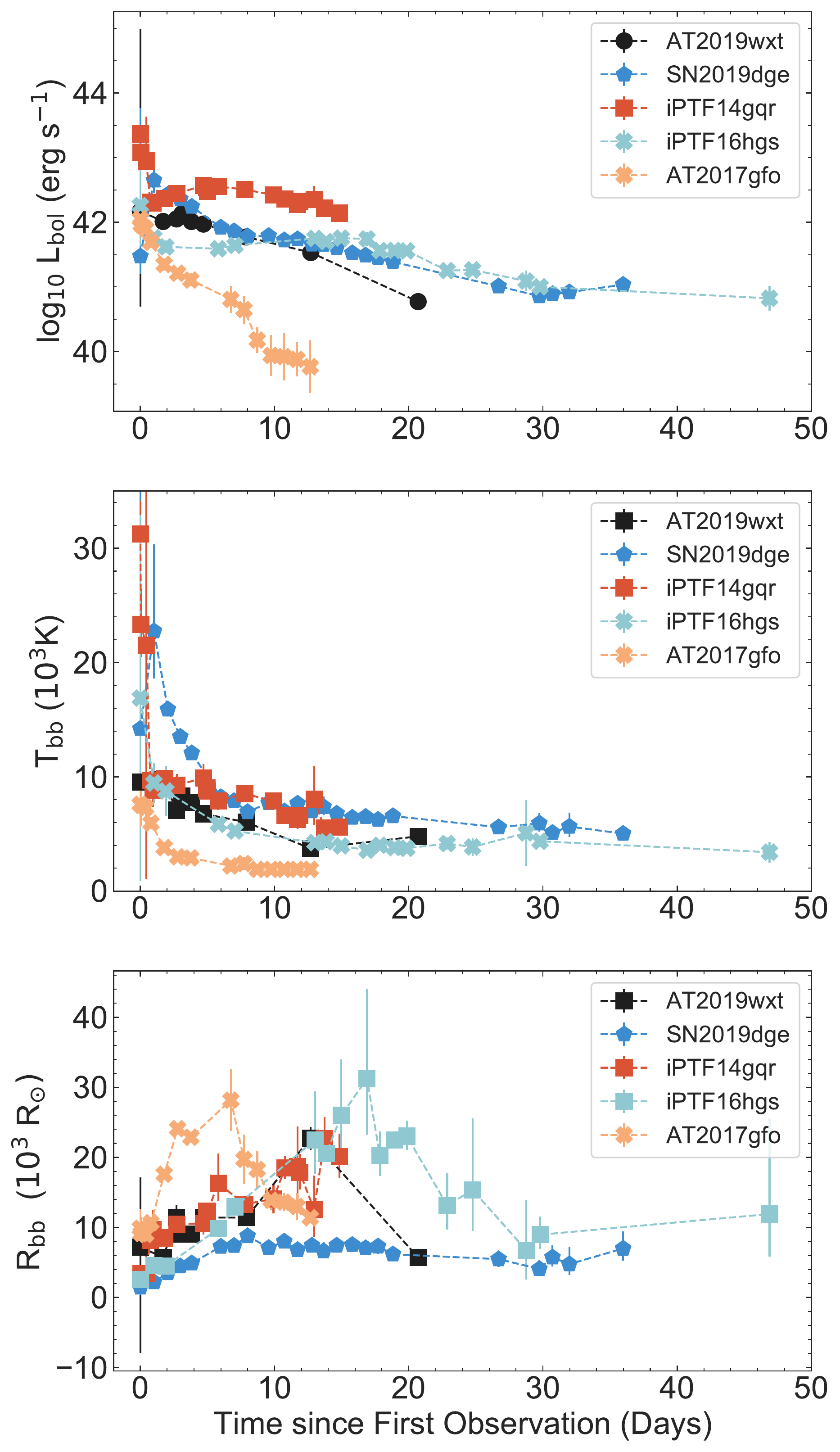}
    \caption{Time evolution of the bolometric luminosity, blackbody temperature, and blackbody radius of AT2019wxt compared with that of the USSN candidates SN2019dge and iPTF14gqr, Ca-rich gap transient iPTF16hgs, and the kilonova AT2017gfo. All times are relative to first optical observation used as a reference epoch for each transient.}
    \label{fig:figure_compare}
    \vspace{-0.5 cm}
\end{figure}

We also compare the evolution of the bolometric luminosity, blackbody temperature, and blackbody radius of \at to that of - i) kilonova AT2017gfo, ii) the USSNe candidates SN2019dge and iPTF14gqr and iii) the Ca-rich gap transient iPTF16hgs in Figure \ref{fig:figure_compare}. As evident from this Figure, the light curve of \at evolves relatively slower than that of AT2017gfo (\citet{gw_ns}; \citet{kn1}; \citet{kn2}; \citet{gfo1}; \citet{cowper_kn}; \citet{kn6}; \citet{kn7};\citet{nicholl2017}; \citet{kn9}; \citet{gfo6}; \citet{kn11}; \citet{kn13}; \citet{kn14}; \citet{kn15}; \citet{kn16}; and \citet{gfo7}).
Following the second peak in the blackbody temperature, T$\textsubscript{bb}$, of \at, the temperature steadily decreases faster than other USSNe candidates but in a manner analogous to iPTF16hgs a Ca-rich ``gap" transient. Moreover, while the photospheric expansion for \at is similar to other USSNe the contraction of the radius evolves on timescales intermediate to the kilonova and Ca-rich gap transients. 

\begin{figure*}[!ht]
    \centering
    \includegraphics[width=0.9\textwidth]{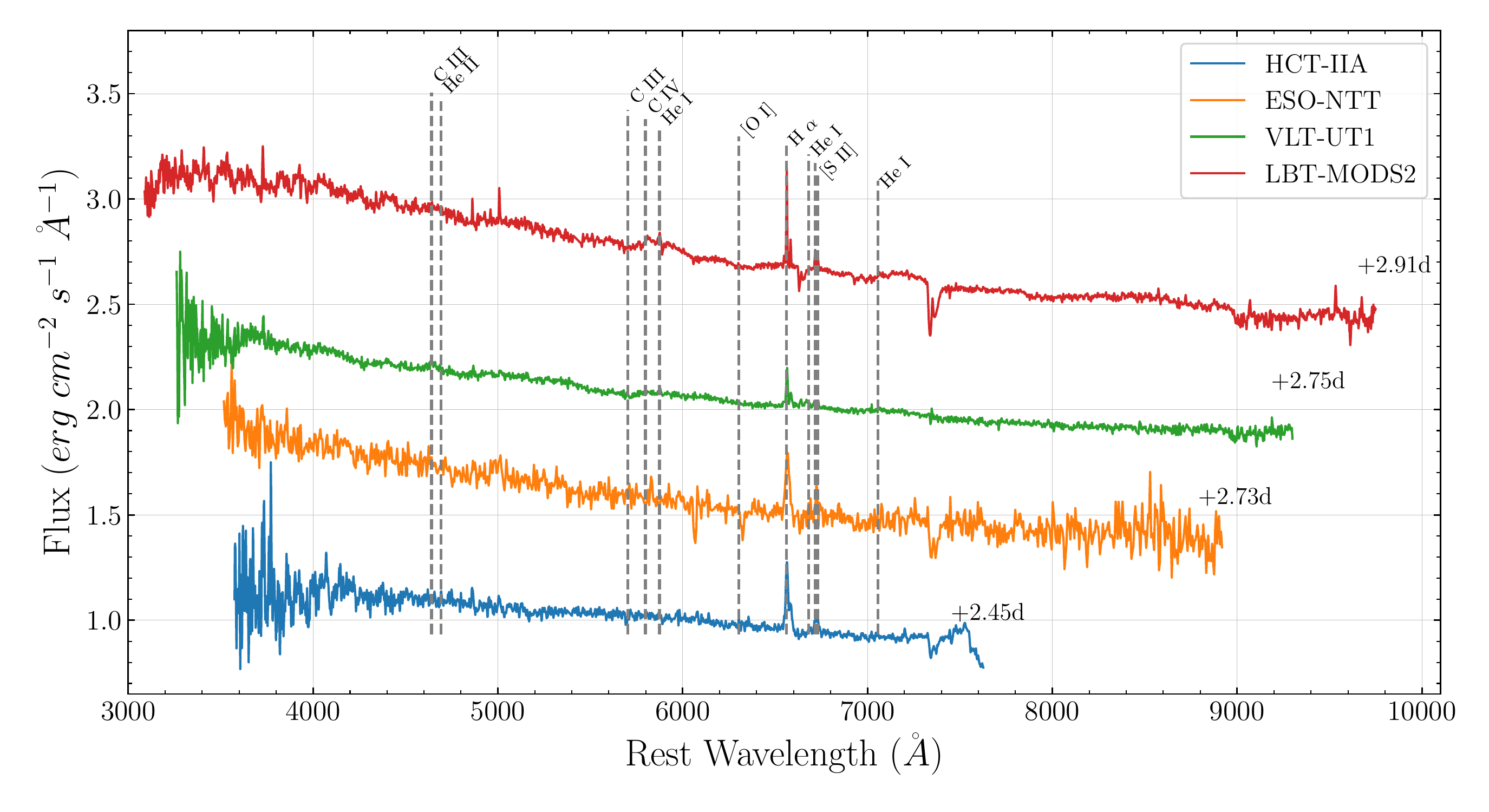}
    \caption{Early-time spectra of AT2019wxt arranged in order of their procurement from bottom to the the top of the plot. These spectra were obtained with various telescopes such as HCT-IIA telescope at IAO-Hanle observatory, VLT and NTT. Each spectrum is normalised and presented with arbitrary offsets for ease of visualisation.}
    \label{fig:spectra}
    \vspace{0.5 cm}
\end{figure*}







\subsection{Spectroscopic Analysis}
We present early-time spectra for \at spanning a net waveband of 3000-10000\,\AA. These spectra were obtained from 2.45-2.75 days post initial peak and during the flux rise towards the second peak. In all of the spectra a blue continuum is observed which we fit with a \texttt{generic\_continuum\_fitting} routine available under the astropy spectroscopy package Specutils\footnote{\url{https://specutils.readthedocs.io/en/stable/index.html}}. In this analysis, the continuum was modelled by smoothing with a median filter to remove the spikes. The source spectra was then normalised by dividing with the fitted continuum. To find significant lines in the spectra we used  the \texttt{find\_line\_derivative} routine in Specutils which finds zero crossings in the spectrum derivative and depending on these, looks for lines above a certain threshold. In our line finding analysis, we set the threshold at 1-$\sigma$ uncertainty in flux spectrum. A summary of significant lines and their presence in different spectra is presented in Table \ref{tab:table_speclines} and Fig. \ref{fig:spectra}. 

\begin{deluxetable}{ lcccc}[!ht]
\renewcommand{\arraystretch}{1.25}
\tablehead{
\colhead{Transition} & \colhead{HCT} &\colhead{ePESSTO+} & \colhead{VLT} & \colhead{LBT}}
\startdata
\hline
He II $\lambda 4686$ & $\checkmark$ & $\checkmark$ & $\checkmark$ & $\checkmark$ \\ 
He I $\lambda 5876$ & $\checkmark$ & $\checkmark$ & $\checkmark$ & $\checkmark$ \\
He I $\lambda 6678$ & $\checkmark$ & $\checkmark$ &  &  \\
He I $\lambda 7065$ & $\checkmark$ & $\checkmark$ & $\checkmark$ & $\checkmark$ \\
H$\alpha ^{*}$ & $\checkmark$ & $\checkmark$ & $\checkmark$ & $\checkmark$ \\
$[$C III$]$ $\lambda 4650$ & $\checkmark$ & $\checkmark$ & $\checkmark$ & $\checkmark$ \\
$[$C III$]$ $\lambda 5696$ & $\checkmark$ & $\checkmark$ &  & \\
$[$C IV$]$ $\lambda 5801$ & $\checkmark$ & $\checkmark$ & $\checkmark$ & $\checkmark$ \\
$[$O I$]$ $\lambda 6300$ & $\checkmark$ & $\checkmark$ &  &  \\
$[$S II$]$ $\lambda 6716 ^{*}$ & $\checkmark$ & $\checkmark$ & $\checkmark$ & $\checkmark$ \\
$[$S II$]$ $\lambda 6731 ^{*}$ & $\checkmark$ & $\checkmark$ & $\checkmark$ & $\checkmark$ \\
\hline
\enddata
\caption{\label{tab:table_speclines} Summary of spectral lines observed in the early-time spectra of \at for various instruments. The transition lines marked with an asterisk indicate host galaxy lines.}
\vspace{-1.5 cm}
\end{deluxetable}

We observe multiple He I lines ($\lambda$5876, $\lambda$6678, $\lambda$7065) and one He II $\lambda$4686 line. Of the He I lines, we notice weakening of the He I $\lambda$6678 as the source flux continues to rise, the He II $\lambda$4686 shows a similar attenuation. This line evolution behaviour is even more noticeable in C III ($\lambda$5696) and O I $\lambda$6300 lines which vanish over a course of 0.02 days between NTT and VLT-U1 observations. Another CIII line at $\lambda$4650 is observed along with a weak C IV $\lambda$5801. Finally, the SII doublet lines - $\lambda$6716 and $\lambda$6731 - from the host galaxy. A prominent H$\alpha$ emission line from the galaxy is also observed.

\subsection{Modeling the Double-Peaked light curve} \label{sec:modeling}
\citet{nakarpiro} show that in case of progenitors that are not enclosed by an extended envelope, only the blue bands exhibit a double-peaked structure in their light curve due to shock-cooling emission (SCE). In contrast, both the blue and red bands show a double-peaked light curve when an extended, low-mass envelope encloses the compact core. Given the red-ward nature of the \textit{i}-band we infer that the double-peak structure observed in \textit{i}-band light curves of \at (see, Fig. \ref{fig:figure_lightcurve}) should indicate a presence of an extended envelope. Moreover, as we show earlier (Sec.\ref{sec:Intro}) light curves of core-collapse SNe which display double peaks in both blue and red bands have been explained in the past with combination of two kinds of emission processes - shock-cooling and radioactive decay of $^{56}$Ni. 

\begin{table}[htp]{
\centering
\begin{tabular}{lp{5cm}c}
\hline\hline
$\theta$ & Description & Prior\\
\hline\hline
logR$\textsubscript{ext}$ & log$_{10}$ of radius of extended envelope (cm) & $\mathcal{U}$(1,18) \\
logM$\textsubscript{ext}$ & log$_{10}$ of mass of extended envelope ($M_{\odot}$) & $\mathcal{U}$(-4,-1) \\
E$\textsubscript{ext,49}$ & Energy in extended envelope divided by 10$^{49}$ erg & $\mathcal{U}$(0.1,20) \\
\hline\hline
\end{tabular}
\caption{\label{tab:table_piropriors} Parameters used for the early-time shock-cooling model and their priors.}}
\vspace{-0.5 cm}
\end{table}

In this section, we follow a similar approach to derive light curve properties from multi-band optical and NIR observations of \at. We explore a scenario such that the early-time emission is dominated by shock-cooling emission from an extended envelope followed by a late-time emission due to radioactive decay of $^{56}$Ni. Hence, we model and subtract this early SCE component before fitting for radioactive decay of $^{56}$Ni. This methodology has been successfully used in the past \citep{yao_dge} to model USSNe with extended envelope. 


We assume that for the early-time emission (t $ < 2 $ days; as seen in \ref{fig:figure_lightcurve}) of \at, the emission is dominated by shock-cooling process. We used the \citet{piro2020} shock-coolig model to constrain the SCE parameters such mass ($M\textsubscript{ext}$), radius ($R\textsubscript{ext}$), and energy ($E\textsubscript{ext, 49}$) of the extended envelope. We fixed the time of the explosion for the model to $t_{exp} \approx -4$\,days based on earliest upper limits available for \at from our \textit{z}-band observation. The SCE parameters were then assigned (wide) flat priors, as presented in Table \ref{tab:table_piropriors}. These priors were informed by low masses and large radii for extended envelopes inferred in previously known transients with early-time SCE. We perform the parameter inference using the emcee package MCMC implementation with a standard Gaussian log-likelihood function and with 100 walkers. The corner plot obtained from this analysis is displayed in Figure \ref{fig:fig_scecorner}, which shows probability distributions for log$_{10}M\textsubscript{ext}$, log$_{10}R\textsubscript{ext}$, and $E\textsubscript{ext, 49}$. We obtain the  best-fitting model for parameter values of $M\textsubscript{ext} = 3.55 ^{+0.12}_{-0.11} \times 10^{-2} \ M_{\odot}$, $R\textsubscript{ext} = 5150.11^{+581.99}_{-513.27} \ R_{\odot}$, and $E\textsubscript{ext, 49} = (10.00^{+6.82}_{-6.75}) \times 10^{49}$ erg. 

\begin{figure}[htp]
    \centering
    \includegraphics[width=\linewidth,  trim={0.5cm 0 0 0}, clip]{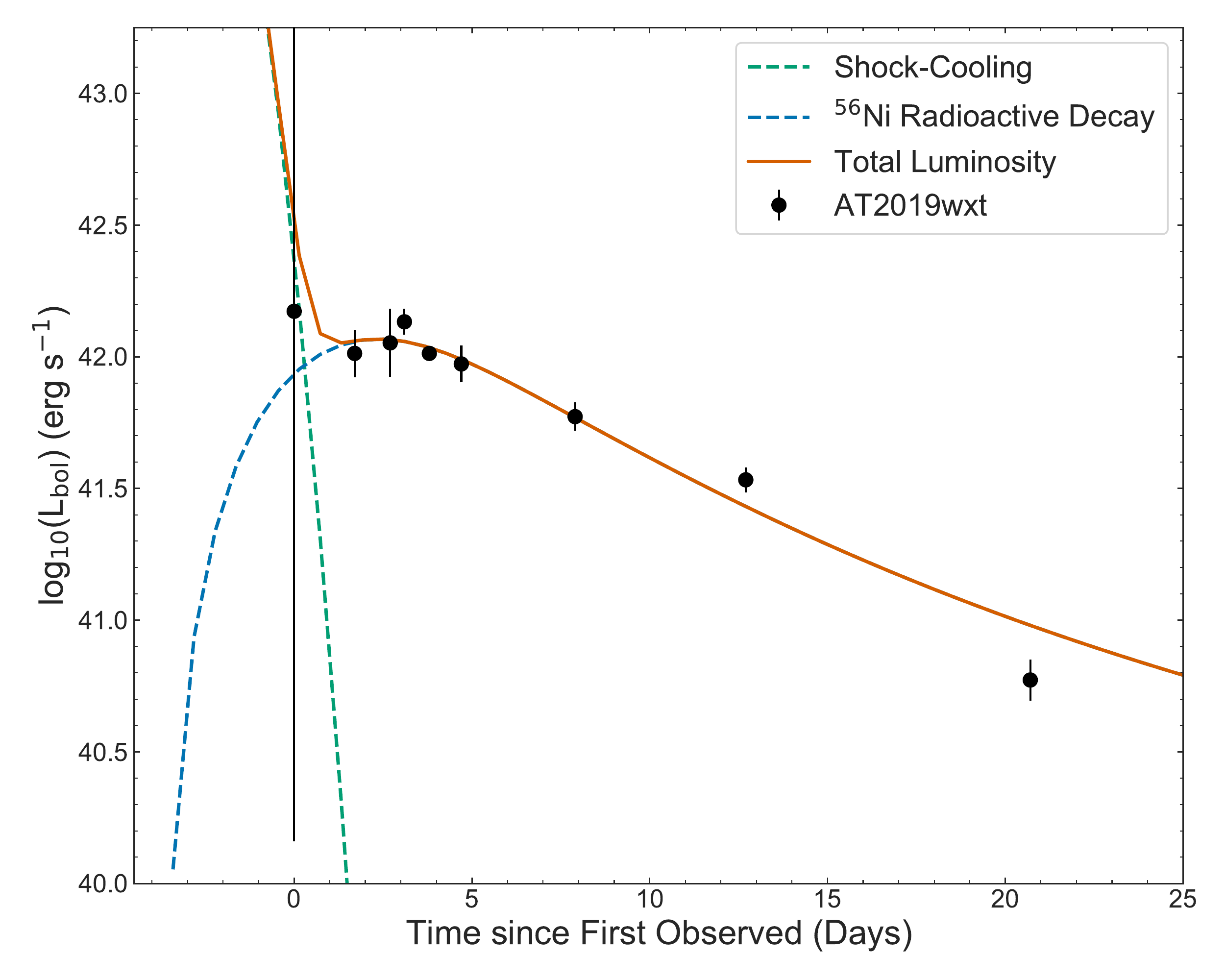}
    \caption{Bolometric light curve of \at plotted with the best-fit shock-cooling model (dashed green line) and the best-fit $^{56}$Ni decay radioactivity model (dashed blue line). The sum of the two models is shown in solid red line.}
    \label{fig:fig_totalmodel}
\end{figure}

We start the late-time emission ($t > 2$ days) fitting procedure by subtracting the best-fit SCE model from the initial light curve to infer the radioactivity-powered properties of \at. We use the $^{56}$Ni decay model by \citet{valenti} to constrain the nickel mass ($M\textsubscript{Ni}$), the characteristic photon diffusion timescale ($\tau_{m}$), and the characteristic $\gamma$-ray timescale ($t_{0}$). The model parameters, are assigned wide, flat priors, as presented in Table \ref{tab:table_arnettpriors}. We again perform the parameter inference using the emcee package MCMC implementation with a standard Gaussian log-likelihood function involving 100 walkers. The corner plot obtained from this MCMC is displayed in Figure \ref{fig:fig_arnettcorner}. We obtain the best-fit radioactivity model for parameter values of M$\textsubscript{Ni} = 2.73^{+0.33}_{-0.18} \times 10^{-2} M_{\odot}$, $\tau_{m} = 4.35^{+1.16}_{-1.43}$ days, and $t_{0} = 12.58^{+1.23}_{-1.34}$ days. The best-fit models for each of the shock-cooling emission and $^{56}$Ni radioactive decay along with total luminosity evolution are presented in the figure \ref{fig:fig_totalmodel}.


\begin{table}[htp]{
\centering
\begin{tabular}{lp{5cm}c}
\hline\hline
$\theta$ & Description & Prior\\
\hline\hline
$\tau_{m}$ & Characteristic photon diffusion time (day) & $\mathcal{U}$(2,6)  \\
logM$\textsubscript{Ni}$ & log$_{10}$ of nickel mass (M$_{\odot}$) & $\mathcal{U}$(-4,-1.25) \\
t$_{0}$ & Characteristic $\gamma$-ray escape time (day) & $\mathcal{U}$(8,100) \\
\hline\hline
\end{tabular}
\caption{\label{tab:table_arnettpriors} Parameters used for the $^{56}$Ni decay radioactivity model and their priors.}}
\end{table}

Based on our parameter estimations, we can calculate the ejecta mass using the an update to the $^{56}$Ni decay model \citep[see Equation 1 in ][]{lyman}:
\begin{equation}\label{eq_ejecta}
    M_{ej} = \left(\frac{\tau_{m}^{2} \ \beta \ c \ v_{ph}}{2 \ \kappa_{opt}}\right)
\end{equation}

where $\beta \simeq 13.8$ is a constant, $\kappa_{opt}\approx{0.07}\,$cm$^2$~g$^{-1}$ is mean optical depth for a SESN, and $v_{ph}$ is photospheric velocity \footnote{We substitute photospheric velocity ($v_{ph}$) in the equation in lieu of the usual ``scale velocity'' given $v_{ph}$ is observationally equivalent to the scale velocity at the maximum bolometric luminosity.}.

In order to approximately calculate the ejecta mass for \at, we assumed that photospheric velocity ($v_{ph}$) should be obtained close to the peak of the bolometric light curve. Therefore, to estimate this velocity we use the evolution of blackbody radius (R$\textsubscript{bb}$) as an approximation for the change of photospheric radius. We then assume a linear expansion of the radius around the second peak (from $t = 1$ day to $t = 15$ days) to calculate the velocity of the ejecta. We find this ejecta velocity to be $\approx{9300}$\,km s$^{-1}$. 

Following from Equation \ref{eq_ejecta} and estimated $v_{ph}$, we find the ejecta mass to be $M_{ej}\approx{0.20}^{+0.12}_{-0.11}M_{\odot}$. This ejecta mass is of the same order of magnitude as the ejecta masses estimated for known USSN objects SN2019dge ($\approx{0.3}M_{\odot}$) and iPTF14gqr ($\approx{0.2}M_{\odot}$), which highlights the USSNe nature of \at. This ejecta mass and velocity translate to \at's kinetic energy being E$\textsubscript{kin,ej}\approx\left({1.01}^{+0.61}_{-0.55}\right)\times 10^{50}$\,erg. We would like to note that due to the assumptions in modelling the emission from \at coupled with scarcity of early-time multi-band observations, there may exist degeneracies between the model parameters, such as between the radius and the energy of the extended envelope \citep{piro2015}.

\subsection{X-ray Image Analysis}
The primary analysis and calibration on X-ray data were performed with version 4.14 of \textit{Chandra}'s CIAO software package \citep{Chandra_Ciao}. The calibration of X-ray data used the database CALDBv4.9.7. We reprocessed the primary and secondary data using the automatic Chandra-\textit{repro} script, resulting in new level 2 event and response files. 

\begin{figure}[!ht]
    \centering
    \includegraphics[width=0.85\columnwidth, trim{0.5cm 0 0 0}, clip]{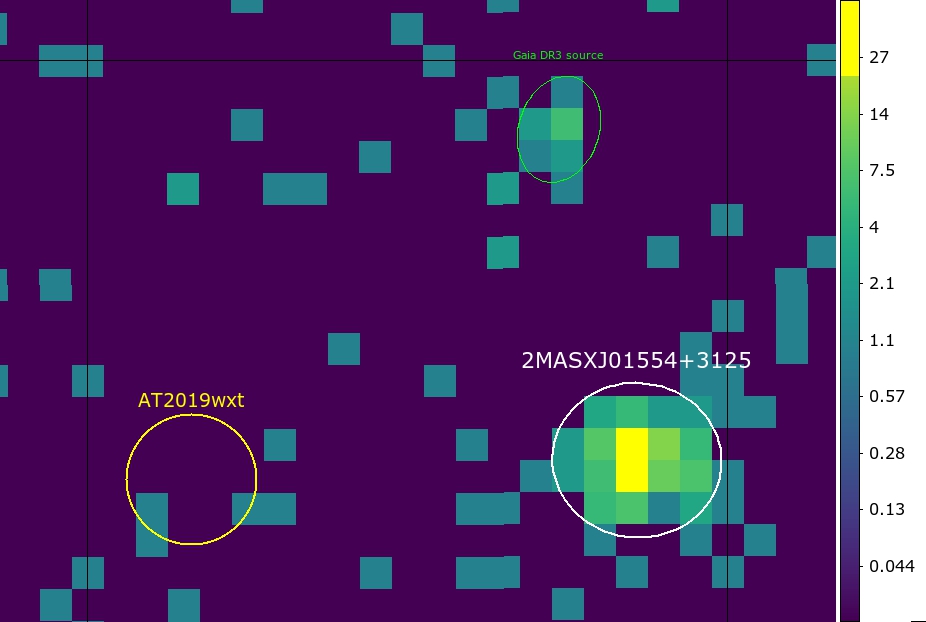}
    \caption{In this panel we present 0.3-10\,keV \textit{Chandra} image of \at during the longest 100ks observation. We plot 1-arcsec radius circle centered optical position of \at. The host galaxy active galactic nucleus (AGN) is indicated by an ellispe of major and minor axes 1.301" and 1.189". We matched the coordinates of host galaxy from Chandra and find a match in both the NED and Gaia catalogs. The image colours are presented in log scale and we do not observe any significant emission at the location of \at, while the host galaxy is confidently detected at the $5\sigma$ level. We also see a faint Gaia transient in the field.}
    \label{fig:chandraimage}
\end{figure}

After this, we obtained X-ray images for the 0.3-8keV energy range and used \textit{wavdetect} to extract all the sources in the region. However, no sources were found in the region of interest and spatially coincident with AT2019wxt. We used the Chandra's \textit{srcflx} routine to arrive at background count rates. This background rate allowed us to establish an upper limit on the source count rate. In this analysis we found the source to be undetectable at 6.51$\times10^{-6}$\,cts/s. We used absorbed power with an index of 2.1 and assumed a neutral Hydrogen density (N$_H$) of $1.8\times10^{20}$\,cm$^{-2}$, to convert this count rate into a 0.3-8 \,keV flux upper limit of $\lesssim 9.93\times10^{-17}$\,erg\,cm$^{-2}$\,s$^{-1}$. Meanwhile, AGN of the host galaxy of AT2019wxt is detected with $5\sigma$ confidence threshold with count rate of 3.52$\times10^{-3}$\,cts\,s$^{-1}$. Using an absorbed power law with index of 1.7, this count rate corresponds to a flux of $5.87\times10^{-14}$\,erg\,cm$^{-2}$\,s$^{-1}$. The X-ray position for the galaxy coincides with the position reported in NED within the uncertainties.   


\subsection{Radio Data Analysis}\label{sec:radio_methods}
\begin{figure}[htp]
    \centering
    \includegraphics[width=\columnwidth]{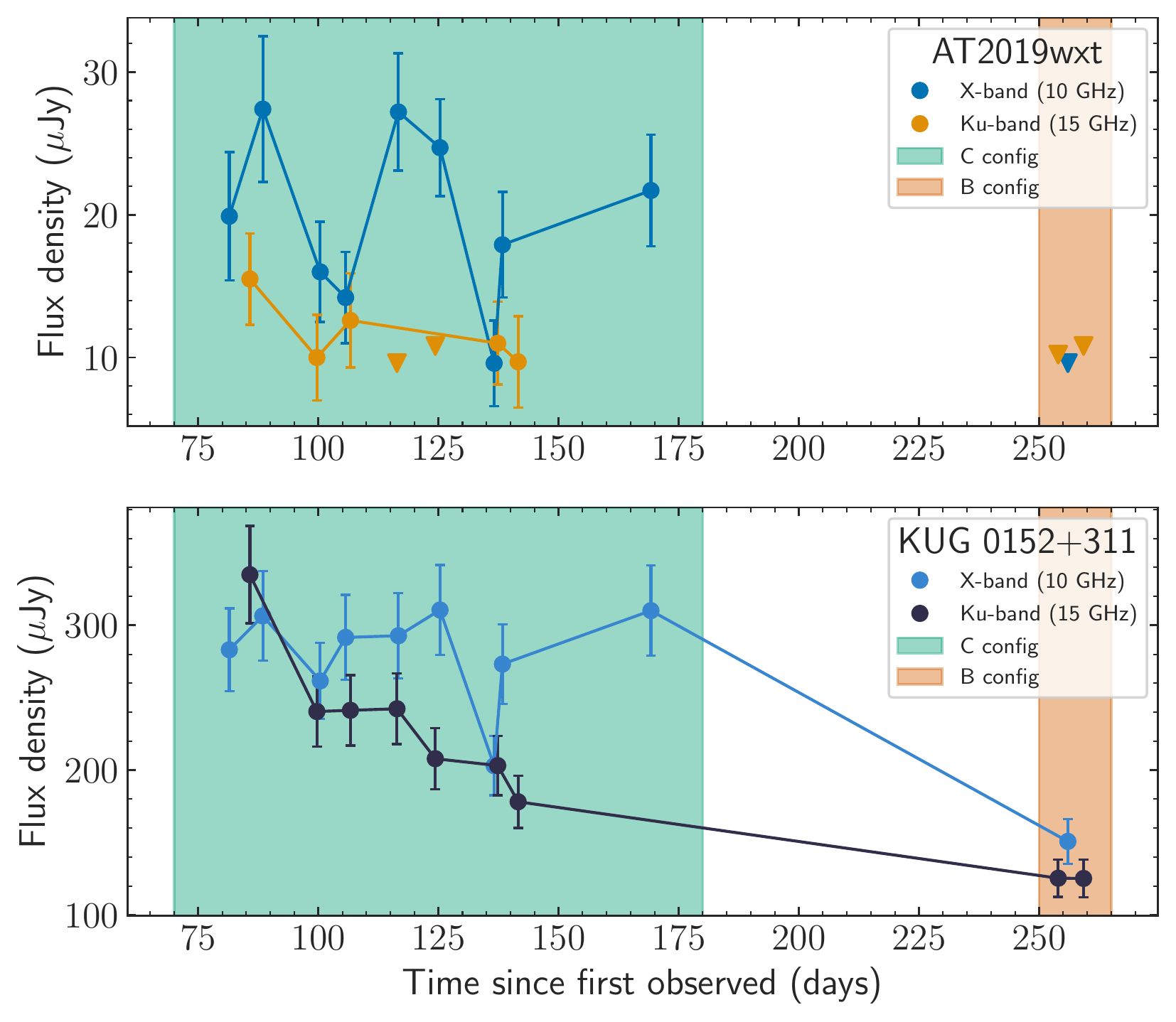}
    \caption{Plot of the radio flux density measurements at the location of \at (top panel) in the X (10\,GHz) and Ku (15\,GHz), compared to the peak flux density measurements of the host galaxy (bottom panel) as a function of time, as reported in Table \ref{tab:radio_obs}. The downward pointing triangles are upper limits (see Section \ref{sec:radio_obs} for more details). 
    }
    \label{fig:radio_lc}
\end{figure}

\begin{figure}
    \centering
    \includegraphics[width=0.48\textwidth, trim={0.5 0 0 0}, clip]{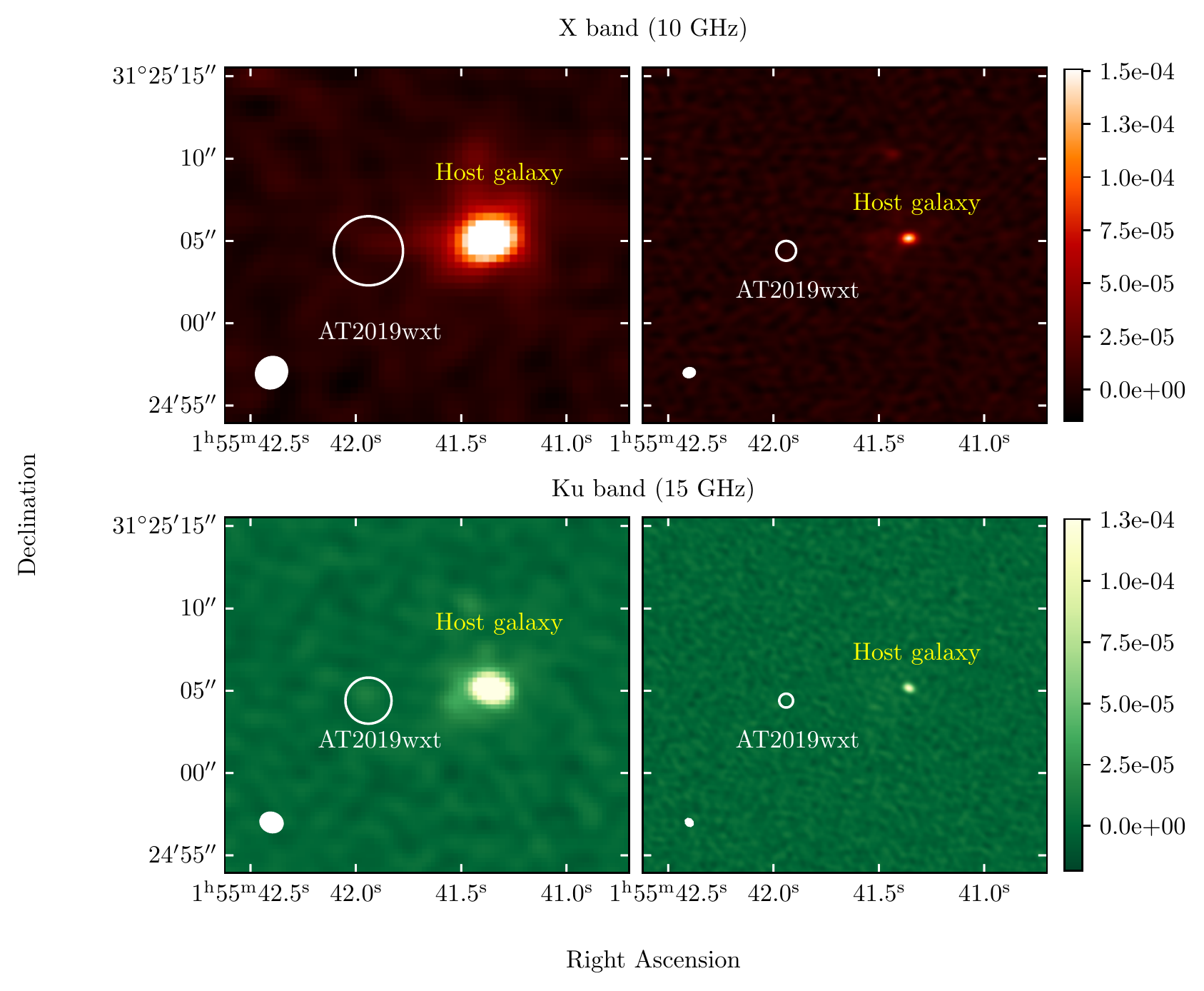}
    \caption{The host galaxy as seen in VLA configurations C (left) and B (right), in the X band (10\,GHz, top) and Ku band (15\,GHz, bottom). The white circles mark the position of AT2019wxt with a radius equal to the FWHM of the nominal VLA synthesized beam of each observation. The synthesised beam ellipse is shown on the left bottom of each panel and the colorbars provide the flux density in units of Jy. See text for discussion.}
    \label{fig:B_C_comp}
\end{figure}

Since all observations in the K band (22\,GHz) were upper limits, we proceed to discuss results from only the X (10\,GHz) and Ku (15\,GHz) bands. Figure \ref{fig:radio_lc} shows the radio flux density measurements at the location of \at (top panel) and the host galaxy, KUG 0152+311 (bottom panel) in the VLA X (10\,GHz) and Ku (15\,GHz) bands (see Table \ref{tab:radio_obs}). 
The host galaxy (KUG 0152+311) is resolved in images taken in X (10\,GHz) and Ku (15\,GHz) bands with the VLA in its C configuration and is marginally resolved in both these bands for the B configuration observations. Moreover, as evident from Figure \ref{fig:B_C_comp}, the host galaxy emission is likely contaminating our measurements at the location of \at in observations taken with the VLA in its more compact C configuration. A strong host galaxy contribution to the flux measured at the \at location is also suggested by the fact that observations taken with the VLA in the more extended B configuration (see colored regions in Figure \ref{fig:radio_lc}) show a drastic decrease in the measured flux density of the host galaxy, and return a non detection at the location of \at (see also Table \ref{tab:radio_obs}). This is compatible with the hypothesis that the more tenuous, extended emission from the host contaminates our C configuration measurements at the location of \at, but goes undetected in the VLA B configuration (see also the right panels in Figure \ref{fig:B_C_comp}).

To test whether the flux we measure at the \at location is dominated by host galaxy emission, we computed  the Pearson correlation coefficient  to check for a potential correlation between such flux and host galaxy flux density measurements \citep[see e.g.][]{2019MNRAS.490.5300D,2019MNRAS.488.5420H}.  We find that the flux measured at the \at location is strongly correlated with that measured for the host galaxy (99.32\% in the X band and 98.71\% in the Ku band). This correlation can be visualised by comparing the time variation of the flux measurements at the AT2019wxt location, and that of the host galaxy (Figure \ref{fig:radio_lc}).  
Next, to quantify the variability in the flux measurements obtained at the \at location and for the \at host galaxy, we use our X and Ku band data and adopt the statistical metrics described in \cite{2015A&C....11...25S}. We briefly introduce them here. We use $N$, F$_{\nu}$ and $\sigma_{\nu}$ to represent the number of observations, flux density measurements and the corresponding errors at frequency $\nu$ (see Table \ref{tab:radio_obs}) respectively. The flux density coefficient of variation can be calculated as:
\begin{equation}
V_\nu = \frac{1}{\overline{F_\nu}} \sqrt{\frac{N}{N-1}\left(\overline{{F_\nu}^2}-\overline{F_\nu}^2\right)}.
\label{eq:V_nu}
\end{equation}
Assuming weights of $w_\nu = 1/\sigma_\nu^2$, the weighted average flux density is calculated as:
\begin{equation}
\xi_{F_\nu} = \frac{\sum_{i=1}^{N} w_{\nu} F_{\nu,i}}{\sum_{i=1}^{N} w_{\nu}}.
\end{equation}
Further, we calculate the reduced-$\chi^{2}$ using the above defined weighted mean flux density
\begin{equation}
\eta_\nu = \frac{1}{N-1} \sum_{i=1}^N \frac{\left(F_{\nu, i} - \xi_{F_\nu} \right)^2}{\sigma_{\nu, i}^2}.
\end{equation}
Using this metric we can calculate the probability for the source to be a variable as:
\begin{equation}
P_{\mathrm{var}} = 1 - \int_{{\eta_\nu}'=\eta_\nu}^{\infty}p\left({\eta_\nu}',N-1\right)\mathrm{d}{\eta_\nu}'.
\end{equation}
where $p(x, n)$ is the $\chi^2$ probability density function for $x$ over $n$
degrees of freedom.
Finally, we also compare the variability metrics computed for the flux measurements (excluding upper limits; see Table \ref{tab:radio_obs}) at the location of \at and the host galaxy, KUG 0152+311, with those that can be obtained from the radio light curves of the well-sampled radio-loud well-sampled Type IIb supernova SN1993J \citep{2007ApJ...671.1959W} at the same timescales as our data. This comparison was made owing to the initial type IIb classification of the event (as discussed in Section \ref{sec:spec_obs}).

\begin{deluxetable}{ccccccc}[!ht]
\tablehead{\multirow{2}{*}{Freq. band} & \multicolumn{2}{c}{AT2019wxt} & \multicolumn{2}{c}{Host} & \multicolumn{2}{c}{SN1993J}\\
         \cline{2-7}
         &$V_\nu$ & $P_{\mathrm{var}}$ (\%) & $V_\nu$ & $P_{\mathrm{var}}$ (\%) & $V_\nu$ & $P_{\mathrm{var}}$ (\%)}
\startdata
\hline
\hline
X (10\,GHz)   & 0.3 & 5.6 & 0.1 & 2.2 & 0.2 & 48.4 \\
Ku (15\,GHz)   & 0.2 & 3.3 & 0.2 & 32.6 & 0.2 & 4.2 $\times 10^{-3}$\\
\enddata
\caption{Radio variability metrics for AT2019wxt compared with SN1993J in X (10\,GHz) and Ku (15\,GHz) bands excluding upper limits. See Section \ref{sec:radio_methods} for discussion.}
\label{tab:radio_var}
\vspace{-1 cm}
\end{deluxetable}
\vspace{-0.5 cm}

As evident from Table \ref{tab:radio_var}, \at and its host galaxy display comparable variability statistics in X-band. The last is much smaller than the variability of SN\,1993J over the same timescales and in the same band. We also note that the same variability analysis for the X band (10\,GHz) observations of GW170817 yields a variability statistic of $P_{\rm{var}}=55.7\%$ over timescales comparable to the ones of our radio observations of \at. Thus, we exclude a  GW170817-like radio counterpart for \at.

The variability statistics in Ku band (15\,GHz) is more complex. A SN\,1993J-like transient would have varied very little in this band over the observed timescales. Hence, the lack of substantial variability in this band would not necessarily indicate a host galaxy origin for the emission measured at the \at location. Moreover, we note that in Ku band the flux from the core of the host varies substantially, and more than the flux measured at the position of \at. 


Overall our results suggest that the radio detections at the location of \at are likely to be related to host galaxy contamination. There is little evidence for the type of variability that would be expected for SN\,1993J-like or GW170817-like events.  

\begin{deluxetable*}{l|cccccccc}[!ht]
\tablecolumns{5} 
\tablewidth{\columnwidth}
\tablehead{\colhead{Transient} & \colhead{Redshift} & \colhead{Host Galaxy} & \colhead{Second Peak} & \colhead{Decay Rate (g-band)} & \colhead{Ejecta Mass} & \colhead{Nickel Mass} & \colhead{Envelope Radius} & \colhead{Envelope Mass}\\
\colhead{Name} & \colhead{} & \colhead{Type} &  \colhead{Magnitude} & \colhead{(mag day$^{-1}$)} & \colhead{M\textsubscript{ej} (M$_{\odot}$)} & \colhead{M\textsubscript{Ni} (10$^{-2} M_{\odot}$)} & \colhead{R\textsubscript{ext}} (10$^{13}$ cm) & \colhead{M\textsubscript{ext}} (10$^{-2} M_{\odot}$) }
\startdata
\hline
AT2019wxt & 0.036 & Compact  & -16.6 & 0.41 & 0.20$^{+0.12}_{-0.11}$ & 2.73$^{+0.33}_{-0.18}$ & 35.8$^{+4.06}_{-3.68}$ & 3.55$^{+0.12}_{-0.11}$\\
SN2019dge & 0.021 & Compact & -15.5 & 0.13 & 0.38$^{+0.02}_{-0.02}$ & 1.57$^{+0.04}_{-0.03}$ & 1.19$^{+0.06}_{-0.05}$ & 9.71$^{+0.28}_{-0.27}$\\
iPTF14gqr & 0.063 & Spiral & -17.5 & 0.21 & 0.24$^{+0.02}_{-0.02}$ & 8.14$^{+0.14}_{-0.15}$ & 6.09$^{+8.73}_{-3.18}$ & 2.59$^{+0.46}_{-0.34}$\\
iPTF16hgs & 0.017 & Spiral & -15.1 & 0.17 & 1.68$^{+0.28}_{-0.25}$ & 2.51$^{+0.20}_{-0.22}$ & 2.45$^{+14.08}_{-1.80}$ & 9.27$^{+3.40}_{-2.48}$\\
\hline
\enddata
\caption{\label{tab:comparetransients} Parameters of different fast-evolving transients - Redshift of the host galaxy, host galaxy type, peak luminosity and decay rate with respect to the second peak of \textit{g}-band observations, ejecta mass $M\textsubscript{ej}$, nickel mass $M\textsubscript{Ni}$, radius and mass of extended envelope $R_{ext}$, $M_{ext}$}
\end{deluxetable*}
\vspace{-1.25 cm}
\section{Discussion and Conclusions}\label{sec:disc}
In this paper, we present optical, near-infrared, radio and X-ray observations and analysis for the peculiar and rapidly-evolving transient \at. The source \at was found on 2019 December 16 UTC 07:19:12 during the search for electromagnetic counterpart to the GW trigger S191213g. At the time it was prime optical counterpart of interest given its rapidly evolving light curve akin to kilonova expected from a binary neutron star merger. The source was intensely followed with \textit{grizyJ} bands photometric and spectroscopic observations for approximately 20.7\,days post initial discovery. The optical light curve shows a prominent double-peaked structure in the \textit{i}-band and a less prominent structure in the \textit{g}-band. The second peak has an absolute magnitude \textit{g}\,$\sim$\,\textit{i}\,$\sim$\,--16.6\,mag. The bolometric light curve derived from the multi-band photometry displayed at least one peak, though with relatively large uncertainties. These uncertainties are a result of lack of multi-band observations of \at in the early-time observations (t$\lesssim$2\,days). We characterise the optical/NIR light curve of \at using a combination of early-time shock-cooling component and a late-time $^{56}$Ni decay component. We estimate that the SCE arises from an extended envelope of mass $\approx 0.035 \ M_{\odot}$  and radius $\approx 35.8 \times 10^{13}$cm. We also estimate that the radioactivity-powered component is composed of nickel mass $\approx 0.027 M_{\odot}$ from which we estimate the ejecta mass $\approx 0.20 M_{\odot}$.

Our long-term radio and X-ray observations of \at spanned a period starting from 3.7 days after initial GW trigger discovery and up to 320 days. Our X-ray observations show no evidence for excess emission at the location of \at. Previously, USSNe have been targeted with \textit{Swift}-XRT observations \citep{de_gqr, yao_dge}, however no X-ray emission was observed for these USSNe with multi-epochal X-ray observations at a flux threshold of  $\lesssim10^{-14}$\,erg\,cm$^{-2}$\,s$^{-1}$. In this work with long-term \textit{Chandra} high-resolution X-ray observations of the source, we obtain one of the most stringent non-detection limits at $\lesssim10^{-17}$\,erg\,cm$^{-2}$\,s$^{-1}$.

On the other hand, while our radio observations show excess emission at the location of \at, there is little evidence for SN\,1993j-like or GW170817-like variability over the timescales of our follow-up. We also discussed the possibility of host galaxy contamination at the location of \at in the radio frequencies (especially at the lowest radio frequencies).

\subsection{Transient Progenitor and Nature of the System}
The double-peaked light curve in the \textit{i}-band indicated presence of an extended envelope around the progenitor of \at. This envelope likely originated from extreme stripping of the outer layers of this progenitor star \citep{nakarpiro}. We find that the radius of the extended envelope inferred from our parameter estimation is an order of magnitude larger than those of previously known USSNe candidates (See Table \ref{tab:comparetransients}). We would like to note that the uncertainties for the radius of the extended envelope are underestimated due to the lack of early-time observations, for which the shock-cooling model was fitted. For rapidly evolving transients such as USSNe, it is extremely critical to obtain observations in more than one bands (specifically, the redder bands) to get a better constraint on luminosity and rise time of the first peak. These early-time constraints can help provide a better estimate for the mass and radius of envelopes surrounding progenitors of USSNe. In addition to early-time multi-band observations, spectra taken during the first peak can help break the degeneracy between the the radius of the envelope and energy of the USSN by providing measurements of the velocity and composition of the extended material \citep{piro2015}.

An estimation of the ejecta mass can provide insight into the process likely causing the stripping of the outer layers of the progenitor. One of the mechanisms that can strip the envelope to this extent is binary mass transfer. Predictions from different stellar evolution models link the ejecta mass to the nature of the progenitor system. Single star evolutionary models of massive stars 
predict relatively high ejecta masses ($M_{ej}>{3}M_{\odot}$) \citep{ss1,ss2,ss3}, much larger than that found for \at (M\textsubscript{ej}$\approx 0.20M_{\odot}$). On the contrary, binary stellar evolution models predict that most Type Ib/c SNe emerged from massive stars in close binary systems, with ejecta mass ranging from $1-5M_{\odot}$ \citep{binary1,binary2}. An even greater degree of stripping can occur in a binary system further reducing ejecta masses to the order of $0.1 M_{\odot}$ and $^{56}$Ni mass of the order of $0.01$ $M_{\odot}$ -- as observed in USSNe \citep{tauris2013}. These USSNe explosions are hence posited to be progenitors of double neutron star systems. The ejecta mass estimated for \at is of the same order of magnitude as that of previously known USSNe candidates. Based on our photometric and spectroscopic analysis, we identify \at as a strong USSNe candidate. While this paper was under development, the ENGRAVE collaboration (Agudo et al., in preparation) independently analysed AT2019wxt and arrived at similar conclusion for source characterisation. 

Spectroscopic observations can provide a clue with respect to the extent of stripping that the progenitor underwent. However, given only two USSNe which are currently known we do not yet have a spectral model for these sources with characteristic features. However, some of the spectral lines that we list in Table \ref{tab:table_speclines} have been previously observed in both SN2019dge and iPTF14gqr sources. In our spectroscopic analysis we see no broad Hydrogen lines indicating a loss of Hydrogen envelope in the system \citep{Filippenko:1997}. 


\at was located in a compact host galaxy, KUG 0152+311, at an offset of 0.5”S, 7.7” E from the galactic center. The USSNe candidate SN2019dge was also located in a compact host galaxy SDSS J173646.73 +503252.3 at a projected offset of 0.5" from the center. Meanwhile, the USSNe candidate iPTF14gqr was located in the outskirts of tidally interacting spiral galaxy IV Zw 155 at a projected offset of 24" from the center. The location of \at at a relatively closer distance to the host galaxy's center matches with the prediction of \citet{tauris2015} that USSNe are found to occur close to their host galaxy's star-forming regions.




\subsection{AT2019wxt and searching for USSNe}
Systematic all-sky surveys such as Zwicky Transient Facility (ZTF; \citealp{ztf}; \citealp{ztf2}) and intermediate Palomar Transient Factory (iPTF; \citealp{iptf1}) have accelerated searches for rapidly evolving transients (e.g. \citealp{ztfeg1}; \citealp{ztfeg2}). We can visualise the progressin transients searches in the last decade based on two key parameters: i) characteristic timescale of the transients- defined as the time taken for the magnitude to change by 0.75 mag from the peak, and ii) peak luminosity. A classification plot based on these quantities called the phase-space diagram \citep[see, Fig. 1][]{lum_time} helps visualise spread between different transients such as SN, USSNe, stripped-envelope SN, kilonovae etc. In this plot, we observe that the transient \at lies in the relatively slower-evolving and greater-luminosity regime compared to the kilonova AT2017gfo. However, post the second-peak it evolves faster compared to the USSNe candidates (SN2019dge and iPTF14gqr) and the Ca-rich gap transient (iPTF16hgs). It also shows black body parameters intermediate to kilonovae and USSNe as seen in Fig. \ref{fig:figure_compare}. We expect upcoming wide-field surveys to open up a new discovery space for faster, fainter transients at large redshifts. A larger sample of transients populating the phase space diagram will be able to outline classes and sub-classes of transients such as SESNe \citep[see, Fig. 18][]{lum_time2}.

The presence of an early-time peak in \at is dominated by SCE. This SCE links to the progenitor stripping and highlights the importance of observing such rapidly-evolving transients in the early stages. One of the main reasons this early-time peak was captured for \at is that the source was coincidentally situated within the LIGO BNS merger region. Hence, global multi-band optical observation campaigns were launched to search for a rapidly-evolving counterpart to the GW trigger S191213g \citep{kasliwal2020}. \citet{Andreoni:2019} highlight the potential of high-cadence observational campaign of the Vera C. Rubin Observatory's Legacy Survey of Space and Time (LSST) \citep{Ivezic:2019} in capturing early SCE emission peak in SNe, which can provide important constraints on the progenitor star. Currently, only two USSNe candidates beyond \at are known among nearly 10000 SNe found thus far. LSST is expected to improve SNe statistics up to a million SNe/yr \citep{LSST:2009}. Scaling to the first order from current ratio of USSNe to the SNe population, we expect $\sim$20 USSNe/yr with LSST at larger redshifts. Early time peak capture of these USSNe will be important in arriving at properties of the source class as highlighted by \at.

\section{Acknowledgements}\label{sec:conc}
J.M.D. and H.S. acknowledge support from the Amsterdam Academic Alliance (AAA) Program, and the European Research Council (ERC) European Union's Horizon 2020 research and innovation program (grant agreement No. 679633; Exo-Atmos). This work is part of the research program VIDI New Frontiers in Exoplanetary Climatology, with project number 614.001.601, which is (partly) financed by the Dutch Research Council (NWO). F.H. and A.J. would like to thank observational support from Chandra X-ray observatory staff. Their work was supported by Chandra observational grant award GO0-21067X. A.B. and A.C. acknowledge support from the National Science Foundation via grant \#1907975. The National Radio Astronomy Observatory is a facility of the National Science Foundation operated under cooperative agreement by Associated Universities, Inc. H.S. would like to thank Yuhan Yao for the methodologies developed to characterise USSNe that we have used in this work and for the insightful discussions throughout the course of this work. Nayana A.J. would like to acknowledge DST-INSPIRE Faculty Fellowship (IFA20-
PH-259) for supporting this research.

%
%
\bibliographystyle{aasjournal} 
\bibliography{sample} 

\appendix
\renewcommand{\thesubsection}{\Alph{subsection}}

\subsection{Observation Tables}
\begin{deluxetable}{lcp{3cm}c}[!ht]
\tabletypesize{\scriptsize}
\tablewidth{\columnwidth}
\tablehead{\colhead{Obs. Time (MJD)} & \colhead{Filter} &\colhead{Magnitude (mag)}& \colhead{Mag error (mag)}}
\startdata
\multicolumn{4}{c}{Pan-STARRS \citep{McBrien:2019}}\\
\hline
58829.348 & z & $>$21.0 & --\\
58830.379 & z & $>$20.3 & --\\
58832.305 & i & $>$19.4 & --\\
58833.305 & i & 19.29 & 0.05\\
58833.320 & i & 19.23 & 0.07\\
58833.335 & i & 19.28 & 0.07\\
\hline
\multicolumn{4}{c}{Palomar P60-inch \citep{gcnfrem}}\\
\hline
58836.687 & g & 19.43 & 0.10\\
58836.703 & g & 19.42 & 0.11\\
58836.684 & r & 19.28 & 0.11\\
58836.700 & r & 19.27 & 0.18\\
58836.690 & i & 19.28 & 0.13\\
58836.706 & i & 19.30 & 0.12\\
\hline
\multicolumn{4}{c}{Lulin 1-m \citep{gcnlulin}}\\
\hline
58836.446 & g & 19.32 & 0.04\\
58836.446 & r & 19.27 & 0.03\\
58836.446 & i & 19.38 & 0.06\\
\hline
\multicolumn{4}{c}{DCT \citep{gcndct}}\\
\hline
58837.126 & g & 19.67 & 0.02\\
58837.114 & r & 19.30 & 0.02\\
58837.120 & i & 19.52 & 0.01\\
58837.126 & z & 19.51 & 0.03\\
\hline
\multicolumn{4}{c}{Pan-STARRS \citep{gcnsmartt}}\\
\hline
58836.434 & g & 19.23 & 0.09\\
58841.211 & g & 20.25 & 0.07\\
58836.436 & r & 19.19 & 0.07\\
58841.213 & r & 20.03 & 0.05\\
58836.438 & i & 19.21 & 0.07\\
58841.214 & i & 19.88 & 0.04\\
58836.439 & z & 19.34 & 0.11\\
58841.216 & z & 19.75 & 0.05\\
58836.441 & y & 19.31 & 0.22\\
58841.218 & y & 19.68 & 0.12\\
\hline
\multicolumn{4}{c}{Wendelstein 2-m \citep{gcnwend}} \\
\hline
58835.000 & g & 19.59 & 0.08\\
58836.000 & g & 19.64 & 0.11\\
58838.000 & g & 19.91 & 0.06\\
58846.000 & g & 22.58 & 0.05\\
58854.000 & g & 23.49 & 0.08\\
58835.000 & i & 20.00 & 0.09\\
58836.000 & i & 19.74 & 0.09\\
58838.000 & i & 19.73 & 0.06\\
58846.000 & i & 20.82 & 0.06\\
58854.000 & i & 22.54 & 0.07\\
58835.000 & J & 20.22 & 0.08\\
58836.000 & J & 19.14 & 0.11\\
58838.000 & J & 18.92 & 0.10\\
58846.000 & J & 19.71 & 0.11\\
58854.000 & J & 20.65 & 0.12
\enddata
\caption{\label{tab:table_photometry}Summary of optical observations of AT2019wxt. The observation of \at at MJD 58833.305 sets the reference epoch for this work.}
\end{deluxetable}

\begin{deluxetable*}{@{\extracolsep{1pt}}llllll}[!htb]
\tablewidth{0.8\columnwidth}
\tabletypesize{\small}
\tablehead{\colhead{Obs Start (MJD)} & \colhead{Telescope} &  \colhead{Instrument}  & \colhead{Wavelength (\AA)} & \colhead{Exposure time (s)} & \colhead{Resolution}}
 \startdata
58835.753 & HCT-IIA & HFOSC2 & 3800-7500 & 3600 & 1200\\
58836.035 & NTT-EPESSTO & EFOSC/1.57 & 3985-9315 & 1200.0061 & 18\AA \\
58836.059 & ESO-VLT-U1 & CCDF-FORS2 & 3400-9600 & 1499.9388 & 10\AA \\
58836.216 & LBT &	MODS2 & 3200-9750  & 3600 & 4\AA \\
\enddata
\caption{\label{tab:specs}Summary of optical spectroscopic observations of \at.}
\vspace{-0.35cm}
\end{deluxetable*}

 \begin{deluxetable}{@{\extracolsep{12pt}}lll}[!ht]
\tablewidth{\columnwidth}
\tabletypesize{\small}
\tablehead{\colhead{Obs Id} & \colhead{Obs Start} & \colhead{Effective}  \\
 \colhead{} &\colhead{Time (MJD)}&\colhead{Exposure (ks)}}
 \startdata
\hline
22458 &	58920.64792	& 49.41\\
23193 &	58922.27917	& 46.45\\
22459 &	59017.46736	& 42.51\\
23283 &	59018.33472	& 49.24\\
22460 &	59077.09375	& 33.45\\
22461 &	59152.85 & 19.82\\
24848 &	59153.8	& 14.89\\
\hline
\enddata
\caption{\label{tab:x_ray}Summary of \textit{Chandra} X-ray imaging observations of \at, over a period of 233 days. All observations were obtained with ACIS-S3 chip in TE mode and VFAINT telemetry format.}
\vspace{-1.25 cm}
\end{deluxetable}

\begin{deluxetable*}{ccccccc}[!ht]
\tablewidth{0.75\textwidth}
\tabletypesize{\footnotesize}
\tablehead{\colhead{Obs. Time} & \colhead{Time Elapsed} &\colhead{VLA config.}& \colhead{VLA band} & \colhead{Obs. Freq.} & \colhead{Source Flux ($\mu$Jy)} &  \colhead{Galaxy Flux ($\mu$Jy)}\\
\colhead{(MJD)} & \colhead{(days)} & \colhead{} & \colhead{} & \colhead{(GHz)} & \colhead{F$_{\nu}\pm\sigma_{\nu}$} & \colhead{ peak F$_{\nu}\pm\sigma_{\nu}$}}
\startdata
\hline
\hline
58837.0 & 3.7 & D & K & 22.0 & $<$ 25.1 & 452 $\pm$ 91  \\
58914.8 & 81.5 & C & X & 10.0 & 19.9 $\pm$ 4.5 & 283 $\pm$ 29 \\
58919.1 & 85.7 & C & Ku & 15.4 & 15.5 $\pm$ 3.2 & 335 $\pm$ 34 \\
58921.7 & 88.4 & C & X & 9.3 & 27.4 $\pm$ 5.1 & 306 $\pm$ 31 \\
58933.0 & 99.7 & C & Ku & 15.4 & 10.0 $\pm$ 3.0 & 240 $\pm$ 24 \\
58933.7 & 100.4 & C & X & 9.8 & 16.0 $\pm$ 3.5 & 262 $\pm$ 26 \\
58939.0 & 105.7 & C & X & 10.0 & 14.2 $\pm$ 3.2 & 292 $\pm$ 29 \\
58940.0 & 106.7 & C & Ku & 15.4 & 12.6 $\pm$ 3.3 & 241 $\pm$ 24 \\
58943.7 & 110.4 & C & K & 22.0 & $<$ 21.6 & 152 $\pm$ 31 \\
58949.7 & 116.4 & C & Ku & 15.4 & $<$ 9.6 & 242 $\pm$ 24 \\
58950.0 & 116.7 & C & X & 9.7 & 27.2 $\pm$ 4.1 & 293 $\pm$ 29 \\
58956.7 & 123.3 & C & K & 22.0 & $<$ 11.7 & 233 $\pm$ 47 \\
58957.6 & 124.3 & C & Ku & 15.3 & $<$ 10.8 & 208 $\pm$ 21 \\
58958.6 & 125.3 & C & X & 10.0 & 24.7 $\pm$ 3.4 & 311 $\pm$ 31 \\
58969.9 & 136.6 & C & X & 10.0 & 9.6 $\pm$ 3.0 & 203 $\pm$ 20 \\
58970.6 & 137.3 & C & Ku & 14.8 & 11.0 $\pm$ 2.9 & 203 $\pm$ 20 \\
58971.6 & 138.3 & C & X & 9.3 & 17.9 $\pm$ 3.7 & 273 $\pm$ 28 \\
58974.9 & 141.6 & C & Ku & 15.7 & 9.7 $\pm$ 3.2 & 178 $\pm$ 18 \\
59002.5 & 169.2 & C & X & 9.4 & 21.7 $\pm$ 3.9 & 310 $\pm$ 31 \\
59087.3 & 254.0 & B & Ku & 15.0 & $<$ 10.2 & 125 $\pm$ 13 \\
59089.3 & 256.0 & B & X & 10.0 & $<$ 9.6 & 151 $\pm$ 15 \\
59092.6 & 259.3 & B & Ku & 15.0 & $<$ 10.8 & 125 $\pm$ 13 \\
\hline
\enddata
\caption{\label{tab:radio_obs}VLA observations of \at along with the integrated flux measurements of the host galaxy KUG\,0152+311. All epochs are with respect to the first optical detection (MJD 58833.305). See Section \ref{sec:radio_obs} for more details.}
\end{deluxetable*}

\clearpage

\subsection{Parameter Estimation}

\begin{figure}[!ht]
    \centering
    \includegraphics[width=\columnwidth]{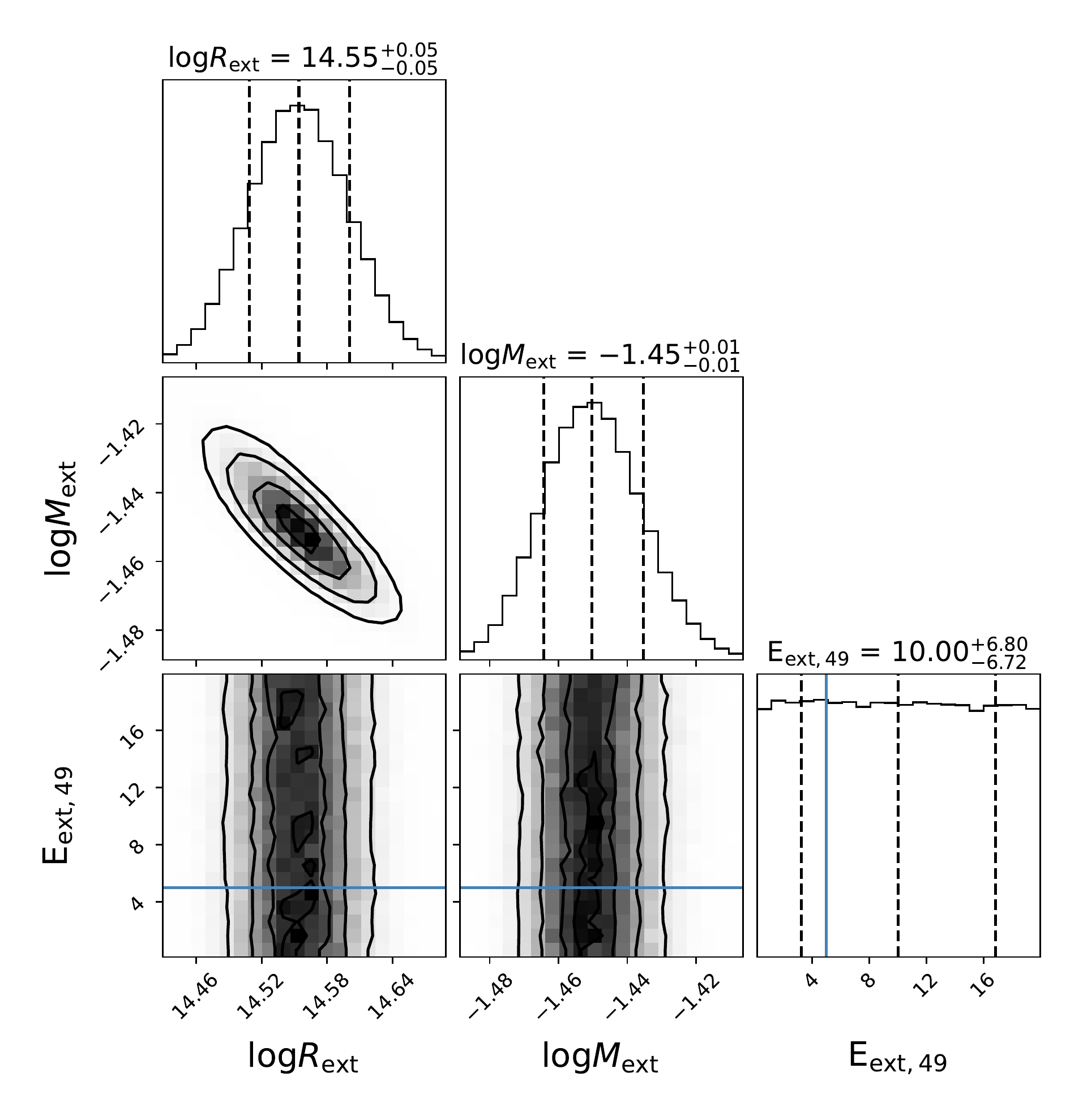}
  \caption{Corner plot obtained from the shock-cooling model displaying the constraints on the posteriors of log$_{10}$ of mass of the envelope (logM$\textsubscript{env}$), log$_{10}$ of radius of envelope (logR$\textsubscript{env}$), and energy of the envelope (E$\textsubscript{ext, 49}$). Along the diagonal, the vertical dashed lines on the histogram indicate the estimate of the best-fit (median) value and the 68\% confidence intervals. The contours represent the 16$^{\mbox{th}}$, 50$^{\mbox{th}}$ and 84$^{\mbox{th}}$ percentile for the respective phase space.}
     \label{fig:fig_scecorner}
\end{figure}

\begin{figure}[!ht]
    \centering
    \includegraphics[width=\columnwidth]{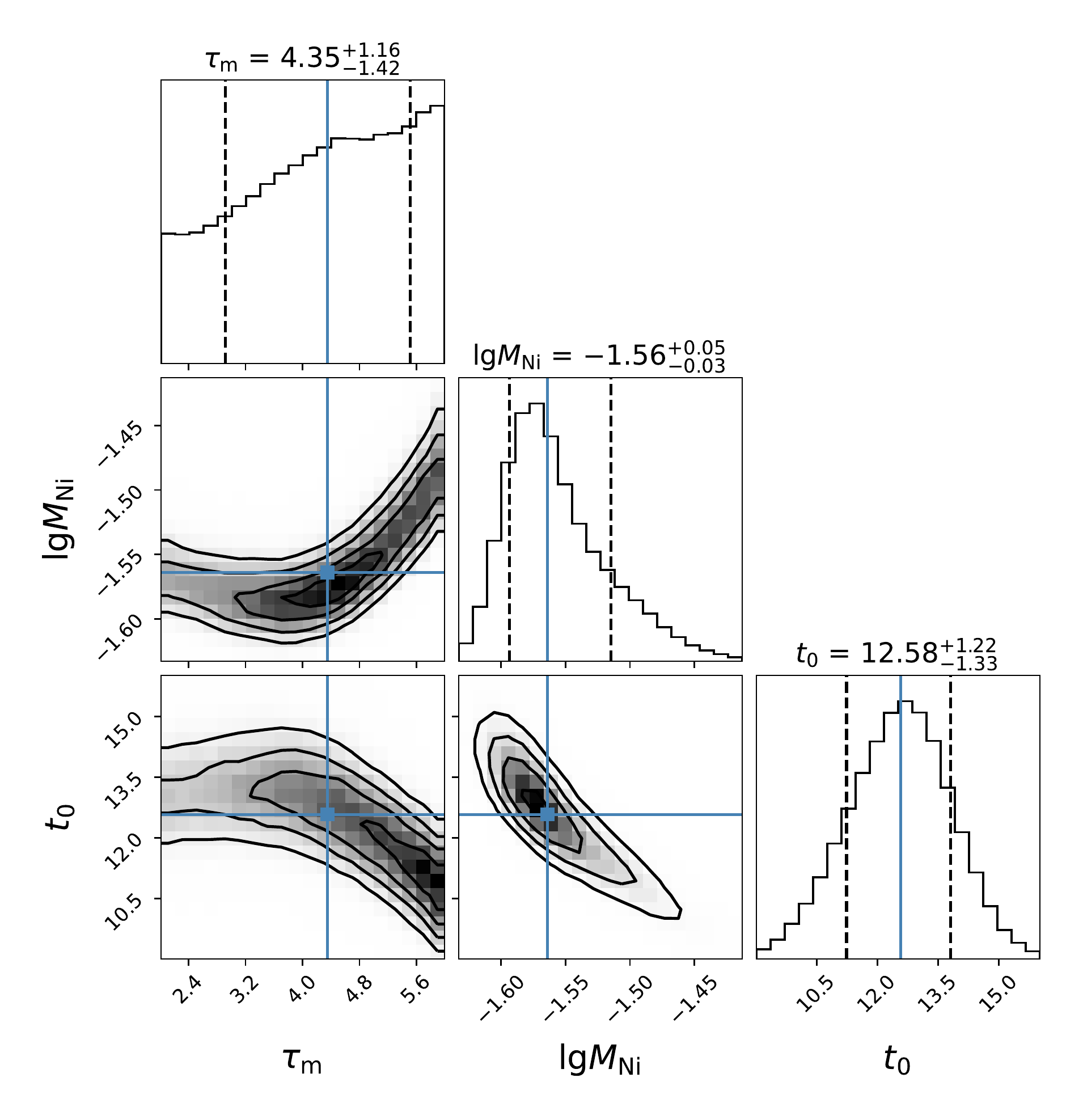}
  \caption{Corner plot obtained from the radioactivity model displaying the constraints on the posteriors of $t_0$, log$_{10}$M$_{Ni}$, and $\tau_{m}$ derived from the radioactivity model. Along the diagonal, the vertical dashed lines on the histogram indicate the estimate of the best-fit (median) value and the 68\% confidence intervals. The contours represent the 16$^{\mbox{th}}$, 50$^{\mbox{th}}$ and 84$^{\mbox{th}}$ percentile for the respective phase space.}
     \label{fig:fig_arnettcorner}
\end{figure}





\end{document}